\documentclass[10pt,aps,prd,reprint,nofootinbib,amsfonts,amssymb,amsmath,preprintnumbers,longbibliography]{revtex4-2}
\pdfoutput=1
\usepackage[utf8]{inputenc}
\usepackage[T1]{fontenc}
\usepackage[british]{babel}
\usepackage[Symbolsmallscale]{upgreek}
\usepackage{mathtools}
\usepackage{isomath}
\usepackage[protrusion=true,tracking=true,kerning=true,spacing=true,final,babel=true]{microtype}
\usepackage{physics}
\usepackage[final]{hyperref}
\usepackage{tikz}

\begin{document}
\title{Primordial black holes from cusp collapse on cosmic strings}

\author{Alexander~C.~Jenkins}
\email{alexander.jenkins@kcl.ac.uk}
\affiliation{Theoretical Particle Physics and Cosmology Group, Physics Department, King's College London, University of London, Strand, London WC2R 2LS, UK}

\author{Mairi~Sakellariadou}
\email{mairi.sakellariadou@kcl.ac.uk}
\affiliation{Theoretical Particle Physics and Cosmology Group, Physics Department, King's College London, University of London, Strand, London WC2R 2LS, UK}

\begin{abstract}
    Primordial black holes (PBHs) are of fundamental interest in cosmology and astrophysics, and have received much attention as a dark matter candidate and as a potential source of gravitational waves.
    One possible PBH formation mechanism is the gravitational collapse of cosmic strings.
    Thus far, the entirety of the literature on PBH production from cosmic strings has focused on the collapse of (quasi)circular cosmic string loops, which make up only a tiny fraction of the cosmic loop population.
    We demonstrate here a novel PBH formation mechanism: the collapse of a small segment of cosmic string in the neighbourhood of a cusp.
    Using the hoop conjecture, we show that collapse is inevitable whenever a cusp appears on a macroscopically-large loop, forming a PBH whose rest mass is smaller than the mass of the loop by a factor of the dimensionless string tension squared, $(G\mu)^2$.
    Since cusps are generic features of cosmic string loops, and do not rely on finely-tuned loop configurations like circular collapse, this implies that cosmic strings produce PBHs in far greater numbers than has previously been recognised.
    The resulting PBHs are highly spinning and boosted to ultrarelativistic velocities; they populate a unique region of the BH mass-spin parameter space, and are therefore a ``smoking gun'' observational signature of cosmic strings.
    We derive new constraints on $G\mu$ from the evaporation of cusp-collapse PBHs, and update existing constraints on $G\mu$ from gravitational-wave searches.
\end{abstract}

\date{\today}
\preprint{KCL-PH-TH/2020-30}

\maketitle

\section{Introduction}
\label{sec:intro}

Primordial black holes (PBHs) have held a prominent place in theoretical cosmology and astrophysics for more than 50 years~\cite{Zeldovich:1967aa,Hawking:1971ei,Carr:1974nx,Carr:1975qj}, playing a wide variety of possible phenomenological r\^{o}les.
Their original motivation was as a source of Hawking radiation~\cite{Hawking:1974sw,Hawking:1974rv,Carr:1976zz}, and this remains an important line of research today~\cite{Arbey:2019vqx,Hooper:2020evu}.
They are natural and well-motivated dark matter (DM) candidates, being massive, nonbaryonic, nonrelativistic, and interacting only through gravity~\cite{Chapline:1975ojl,Carr:2016drx,Carr:2020xqk}.
Binary PBHs are interesting potential sources of gravitational waves (GWs)~\cite{Bird:2016dcv,Clesse:2016vqa,Sasaki:2016jop,Raidal:2017mfl,Sasaki:2018dmp,Raidal:2018bbj,DeLuca:2020qqa}, and are a possible formation channel for the unexpectedly massive BH binaries observed by Advanced LIGO and Advanced Virgo~\cite{Harry:2010zz,TheLIGOScientific:2014jea,TheVirgo:2014hva,Aasi:2013wya} in their first two observing runs (O1 and O2)~\cite{TheLIGOScientific:2016htt,TheLIGOScientific:2016pea,LIGOScientific:2018mvr}.
PBHs could also act as the seeds for the formation of cosmic structures~\cite{Afshordi:2003zb,Carr:2018rid}, particularly the supermassive black holes (SMBHs) observed at the centres of most galaxies~\cite{Silk:1997xw,Bean:2002kx,Clesse:2015wea}.

The most commonly-invoked mechanism for PBH formation is the collapse of large overdensities in the early Universe.
However, this is only possible if the primordial power spectrum has a very large amplitude at small scales, which typically requires a certain degree of inflationary model building and fine tuning\footnote{%
See however Ref.~\cite{Carr:2019kxo} for a recent exception.}
---either in terms of the inflationary field content~\cite{GarciaBellido:1996qt}, or in terms of adding features to the inflaton potential~\cite{Ivanov:1994pa}---and is subject to constraints from CMB measurements of the power spectrum at large scales~\cite{Byrnes:2018txb,Carrilho:2019oqg} by the Planck satellite~\cite{Planck:2006aa,Akrami:2018vks,Akrami:2018odb}.
These constraints become much stronger in the presence of primordial non-Gaussianity, as PBH formation then sources large isocurvature modes~\cite{Chisholm:2005vm,Tada:2015noa,Young:2015kda} which are ruled out by Planck~\cite{Akrami:2018odb}.
Given a generic inflationary theory, there is therefore no guarantee of PBH formation.
It is thus of great interest to find alternative PBH formation mechanisms that are as generic as possible.

One such alternative is the gravitational collapse of \emph{cosmic strings}~\cite{Kibble:1976sj,Vilenkin:1984ib,Hindmarsh:1994re,Vilenkin:2000jqa}: 1+1-dimensional topological defects which are generic predictions of many extensions to the Standard Model~\cite{Jeannerot:2003qv}.
On macroscopic scales, cosmic strings are effectively described by a single parameter---their tension $\mu$, which is conventionally written in the dimensionless combination $G\mu$, and is linked to the energy scale $\eta$ at which the cosmic strings are formed by the relation $G\mu\sim(\eta/m_\mathrm{P})^2\ll1$, where $m_\mathrm{P}$ is the Planck mass.
This tension characterises the gravitational influence of the strings, and is subject to constraints of order $G\mu\lesssim10^{-7}$ from CMB observations~\cite{Kaiser:1984iv,Ade:2013xla,McEwen:2016aob} and of order $G\mu\lesssim10^{-11}$ from GW searches~\cite{Shannon:2015ect,Lasky:2015lej,Verbiest:2016vem,Blanco-Pillado:2017rnf,Abbott:2017mem,LIGOScientific:2019vic,Auclair:2019wcv}.
Hawking first showed in Ref.~\cite{Hawking:1987bn} that PBH formation is the inevitable endpoint of the evolution of circular cosmic string loops,\footnote{%
Strictly speaking this is untrue for loops of length $\ell\lesssim\ell_\mathrm{P}(G\mu)^{-3/2}\approx10^{-18}\,\mathrm{m}\times(G\mu/10^{-11})^{-3/2}$ (where $\ell_\mathrm{P}$ is the Planck length), which unwind and disperse before they become compact enough to form a PBH~\cite{Helfer:2018qgv,Aurrekoetxea:2020tuw}.
However, we are interested in macroscopically-large loops, which lie well above this dispersion regime for any value of $G\mu$ consistent with observational constraints.}
and PBH formation from circular loop collapse has remained an active research topic ever since~\cite{Polnarev:1988dh,Hawking:1990tx,Garriga:1992nm,Caldwell:1993kv,Garriga:1993gj,Caldwell:1995fu,MacGibbon:1997pu,Helfer:2018qgv,James-Turner:2019ssu,Aurrekoetxea:2020tuw}.
However, (quasi)circular collapse is only possible if all three components of the loop's angular momentum are smaller than those of a typical loop by a factor of $\sim G\mu$~\cite{Vilenkin:2000jqa}.
This mechanism is thus finely-tuned, and only a very small fraction of the cosmic loop population is expected to collapse in this way.

In this \emph{Article}, we show that circular loop collapse is not the dominant mechanism for PBH formation from cosmic strings.
We demonstrate, using the hoop conjecture~\cite{Thorne:1972ji,Misner:1974qy}, that cusps on cosmic string loops must inevitably collapse to form PBHs.
Since cusps are generic features of cosmic string loops~\cite{Turok:1984cn,Vilenkin:1984ib,Hindmarsh:1994re,Vilenkin:2000jqa}, and do not rely on finely-tuned loop configurations like circular collapse, this implies that the rate of PBH formation from cosmic strings has been drastically underestimated in the prior literature.

The remainder of this \emph{Article} is structured as follows.
In Sec.~\ref{sec:nambu-goto} we recall the flat-space equations of motion for cosmic string loops in the Nambu-Goto approximation.
In Sec.~\ref{sec:hoop-conjecture} we describe the hoop conjecture, and show that it implies PBH formation on cosmic string loops whenever part of the loop passes a given velocity threshold.
In Sec.~\ref{sec:cusp-collapse} we show that cusps satisfy this condition for the hoop conjecture, and must therefore form PBHs.
In Sec.~\ref{sec:pbh-properties} we estimate the properties of the resulting PBHs, and show that they are highly spinning and boosted to ultrarelativistic velocities.
In Sec.~\ref{sec:pseudocusp-collapse} we make our results even more generic by considering pseudocusps, and showing that they too collapse if they pass a certain threshold.
In Sec.~\ref{sec:backreaction} we argue that gravitational backreaction acts too slowly to prevent PBH formation.
In Sec.~\ref{sec:radiation} we discuss the radiation of mass, linear momentum, and angular momentum from the collapse.
In Sec.~\ref{sec:loop-pbh-dynamics} we explore the qualitative behaviour of the loop-PBH system following the collapse.
In Sec.~\ref{sec:mass-spectrum} we calculate the PBH mass spectrum resulting from a network of cosmic string loops, and derive stringent new constraints on $G\mu$ from PBH evaporation.
In Sec.~\ref{sec:populations} we argue that cusp-collapse PBHs inhabit a unique region of the BH mass-spin parameter space.
In Sec.~\ref{sec:gws} we estimate the GW emission from cusp collapse, calculate the corresponding stochastic GW energy spectrum, and derive updated constraints on $G\mu$ from LIGO/Virgo observations and from Pulsar Timing Arrays (PTAs), as well as updated forecasts for LISA~\cite{Audley:2017drz}.
Finally, in Sec.~\ref{sec:summary} we summarise our results.
Throughout, we use units with $c=1$ and $G,\hbar\ne1$, and use the relativist's metric signature $(-,+,+,+)$.

\section{Nambu-Goto loop dynamics}
\label{sec:nambu-goto}

When studying the gravitational effects of cosmic strings, it is usually convenient to use the Nambu-Goto approximation, in which the string is treated as a classical object with zero thickness.
This is a good approximation for string loops larger than a critical size,
    \begin{equation}
    \label{eq:ell-star}
        \ell_*\equiv\frac{\delta}{G\mu}\approx\frac{\ell_\mathrm{P}}{(G\mu)^{3/2}}\approx5.1\times10^{-19}\,\mathrm{m}\,\times\qty(\frac{G\mu}{10^{-11}})^{-3/2},
    \end{equation}
    where $\delta\approx(\mu/\hbar)^{-1/2}$ is the string width, and $\ell_\mathrm{P}$ is the Planck length.
Loops smaller than $\ell_*$ rapidly lose their energy through particle radiation~\cite{Srednicki:1986xg,Hindmarsh:1994re} or through topological unwinding and dispersion~\cite{Helfer:2018qgv,Aurrekoetxea:2020tuw}, and are therefore uninteresting for our purposes.

We consider the dynamics of Nambu-Goto cosmic strings in flat spacetime, which is a good approximation on scales much smaller than the cosmological horizon and much larger than the string width.
The string's trajectory traces out a 1+1-dimensional surface called the worldsheet.
This is parameterised by the coordinates $(\tau,\sigma)$, with its embedding in the flat background given by $X^\mu(\tau,\sigma)$.
For a closed loop, $\sigma$ is periodic in some interval $[0,\ell)$.

We eliminate the gauge freedom of the worldsheet coordinates by choosing $\tau$ such that $\tau=t=X^0$, and choosing $\sigma$ such that the worldsheet is conformally flat.
We further choose our coordinates in the external spacetime such that the loop's centre of mass is stationary.
The equations of motion for the loop's position $\vb*X(t,\sigma)$ are then~\cite{Kibble:1982cb,Turok:1984cn,Vilenkin:2000jqa}
    \begin{equation}
    \label{eq:EoM}
        \ddot{\vb*X}=\vb*X'',\qquad|\dot{\vb*X}|^2+|\vb*X'|^2=1,\qquad\dot{\vb*X}\vdot\vb*X'=0,
    \end{equation}
    where dots and primes denote derivatives with respect to $t$ and $\sigma$, respectively.
We can decompose the solution into left- and right-moving modes,
    \begin{equation}
        \vb*X(t,\sigma)=\frac{1}{2}\qty[\vb*X_+(\sigma_+)+\vb*X_-(\sigma_-)],\qquad\sigma_\pm\equiv t\pm\sigma.
    \end{equation}
Imposing Eq.~\eqref{eq:EoM} then gives the constraint
    \begin{equation}
    \label{eq:left-right-constraint}
        |\dot{\vb*X}_+|=|\dot{\vb*X}_-|=1.
    \end{equation}
The resulting loop solutions oscillate with period $\ell/2$, with their coordinate length $\int_0^\ell\dd{\sigma}|\vb*X'|$ oscillating over time.
However, $\ell$ itself remains constant (neglecting for now gravitational radiation from the loop), and is the length the loop would have if it were stationary---we refer to it as the ``invariant length.''

The energy-momentum tensor of the loop is derived from the Nambu-Goto action, and is given by an integral over $\sigma$,
    \begin{equation}
    \label{eq:energy-momentum}
        T^{\mu\nu}(t,\vb*x)=\mu\int_0^\ell\dd{\sigma}(\dot{X}^\mu\dot{X}^\nu-X'^\mu X'^\nu)\delta^{(3)}[\vb*x-\vb*X(t,\sigma)].
    \end{equation}
In particular, its mass density is
    \begin{equation}
    \label{eq:T00}
        T_{00}(x)=\mu\int_0^\ell\dd{\sigma}\delta^{(3)}[\vb*x-\vb*X(t,\sigma)],
    \end{equation}
    which shows that the loop's total mass is constant, and is equal to the invariant length times the tension,
    \begin{equation}
        M_\mathrm{loop}=\int_{\Sigma_t}\dd[3]{\vb*x}T_{00}=\mu\ell,
    \end{equation}
    where $\Sigma_t$ is any hypersurface of constant $t$.

\section{The hoop conjecture}
\label{sec:hoop-conjecture}

The hoop conjecture, first formulated by Thorne in Ref.~\cite{Thorne:1972ji}, is a powerful diagnostic for the formation of BH horizons, which circumvents the need to solve the full nonlinear Einstein equation.
The conjecture states that ``horizons form when, and only when, a mass $M$ gets compacted into a region whose circumference in every direction is $\mathcal{C}\le4\uppi GM$''~\cite{Misner:1974qy}.
In other words, if a sphere containing mass $M$ fits inside its own Schwarzschild radius $r_\mathrm{S}\equiv2GM$, it must form a black hole.
This conjecture is intentionally somewhat vaguely defined; in particular, there is no unambiguous way to assign a mass to the gravitational field inside the sphere.\footnote{%
Doing so would require a \emph{quasi-local} measure of gravitational mass in GR, which is a notoriously difficult open problem. See Ref.~\cite{Szabados:2009eka} for a review.}
For the present situation, we include only the mass due to the matter fields,
    \begin{equation}
    \label{eq:M-sphere}
        M_\mathrm{sphere}\equiv\int_{\mathcal{B}_r}\dd[3]{\vb*x}T_{00}(t,\vb*x),
    \end{equation}
    where $\mathcal{B}_r$ is a ball of radius $r$, and the mass is a function of time and of the centre of the ball.\footnote{%
We emphasise that including only the mass due to the matter fields is a conservative choice, as neglecting the mass due to the gravitational field means that we are underestimating the ``total'' quasi-local mass inside the sphere, however that is to be defined.
As such, we must sacrifice the ``only when'' part of the hoop conjecture, but this is inessential for our purposes here.}
The hoop conjecture then predicts BH formation if
    \begin{equation}
    \label{eq:hoop-condition}
        \frac{2GM_\mathrm{sphere}}{r}\ge1.
    \end{equation}
    which we refer to as the ``hoop condition.''

\begin{figure}[t!]
    \includegraphics[width=0.48\textwidth]{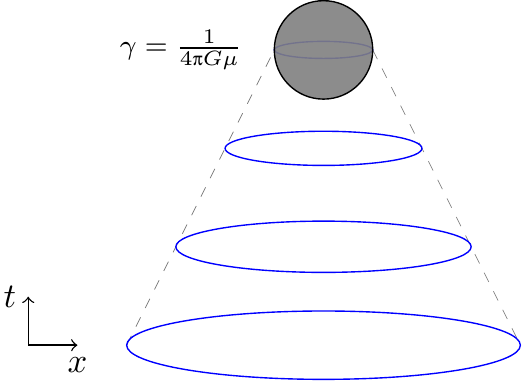}
    \caption{%
    A collapsing circular cosmic string loop (in blue) forms a PBH (in grey) once its Lorentz factor satisfies Eq.~\eqref{eq:gamma-condition-circular}.}
    \label{fig:circular-collapse}
\end{figure}

We are interested in cosmic string loops which lead to PBH formation, i.e., solutions to Eq.~\eqref{eq:EoM} which satisfy the hoop condition~\eqref{eq:hoop-condition} at some point in their evolution.
The simplest example is a circular loop, which contracts at an accelerating rate until the entire loop is compact enough to form a PBH, as illustrated in Fig.~\ref{fig:circular-collapse}.
This occurs within a single loop oscillation period, and results in a PBH of mass $M_\mathrm{PBH}\sim M_\mathrm{loop}=\mu\ell$, which is smaller than the original loop by a factor of $G\mu$.
One can show that the horizon forms only once the loop's Lorentz factor satisfies
    \begin{equation}
    \label{eq:gamma-condition-circular}
        \gamma=\frac{1}{4\uppi G\mu}.
    \end{equation}
For observationally-relevant values of $G\mu$, this corresponds to an ultrarelativistic contraction velocity $v\simeq1-8\uppi^2(G\mu)^2$, and we can understand the PBH formation as being due to relativistic length contraction.
This mechanism for PBH formation from (quasi)circular loops has been studied extensively in the literature~\cite{Hawking:1987bn,Polnarev:1988dh,Hawking:1990tx,Garriga:1992nm,Caldwell:1993kv,Garriga:1993gj,Caldwell:1995fu,MacGibbon:1997pu,Helfer:2018qgv,James-Turner:2019ssu,Aurrekoetxea:2020tuw}.
However, circular collapse is only possible if all three components of the loop's angular momentum are smaller than those of a typical loop by a factor of $\sim G\mu$~\cite{Vilenkin:2000jqa}.
Circular collapse is thus finely-tuned, and only a very small fraction of the cosmic loop population are expected to collapse in this way.

This naturally leads one to ask whether generic (i.e. noncircular) loops can form PBHs.
It is easy to see intuitively that a Lorentz factor of order $\gamma\sim(G\mu)^{-1}$ like that in Eq.~\eqref{eq:gamma-condition-circular} is a necessary condition for PBH formation, even for noncircular loops.
Suppose we want to form a PBH which contains some fraction $f$ of the loop's mass, $M_\mathrm{PBH}=f\mu\ell$, corresponding to a $\sigma$ interval of $\upDelta\sigma=f\ell$.
This length of string must be compacted into a region of diameter $\lesssim2r_\mathrm{S}=4GM_\mathrm{PBH}$.
The ratio between this lengthscale and the corresponding $\sigma$ interval is related to the loop's tangent vector,
    \begin{equation}
        |\vb*X'|\equiv\dv{|\vb*X|}{\sigma}\approx\frac{\upDelta|\vb*X|}{\upDelta\sigma}\lesssim\frac{4GM_\mathrm{PBH}}{f\ell}=4G\mu.
    \end{equation}
We can relate the tangent vector to the loop's dynamics by rearranging the second equation in~\eqref{eq:EoM} to get
    \begin{equation}
    \label{eq:tangent-vector-gamma}
        |\vb*X'|=\sqrt{1-|\dot{\vb*X}|^2}=\frac{1}{\gamma},
    \end{equation}
    which shows that the hoop condition is generically satisfied if part of the loop has a large enough Lorentz factor,
    \begin{equation}
    \label{eq:gamma-condition}
        \gamma\gtrsim\frac{1}{4G\mu}.
    \end{equation}
This is not a sharp bound, just an order-of-magnitude estimate.
The corresponding (exact) inequality~\eqref{eq:gamma-condition-circular} for circular loops agrees to within a factor of $\uppi$.
We expect Eq.~\eqref{eq:gamma-condition} to have a similar level of accuracy for generic loop configurations.

Note that in the above argument we have \emph{not} assumed that the entire loop must be moving at such high velocities, only some fraction $f$ of it.
This is in contrast with the literature on (quasi)circular loop collapse, which has looked exclusively at cases where \emph{all} of the loop's mass ends up behind the PBH horizon.\footnote{%
In fact, Ref.~\cite{Polnarev:1988dh} briefly mentions the possibility of forming a PBH from just part of the loop, but does not discuss this idea in any detail.}
The argument sketched above therefore suggests a change in focus: rather than looking at PBH formation \emph{from} loops, we should be concerned with PBH formation \emph{on} loops.

\section{PBHs from cusp collapse}
\label{sec:cusp-collapse}

\begin{figure}[t!]
    \includegraphics[width=0.48\textwidth]{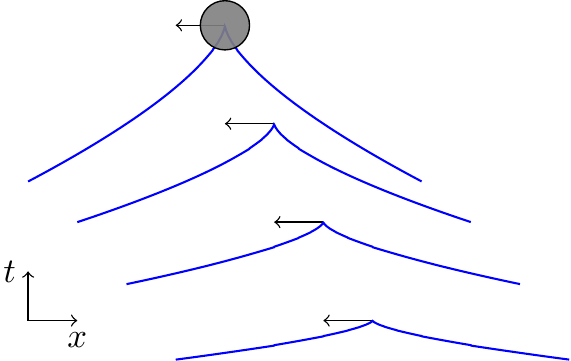}
    \caption{%
    A segment of a cosmic string loop (in blue) becomes more compact as it develops a cusp.
    Once it satisfies the hoop condition~\eqref{eq:hoop-condition} it collapses to form a PBH (in grey).}
    \label{fig:cusp-collapse}
\end{figure}

Solutions to Eq.~\eqref{eq:EoM} generically contain \emph{cusps}: points on the loop which instantaneously reach the speed of light, $|\dot{\vb*X}|=1$~\cite{Turok:1984cn,Vilenkin:1984ib,Hindmarsh:1994re,Vilenkin:2000jqa}.
The Lorentz factor at a cusp diverges, instantaneously compacting a finite fraction of the loop's mass into an infinitesimally small region.
In light of Eq.~\eqref{eq:gamma-condition}, it is clear that cusps should therefore generically lead to some fraction of the loop's mass being enclosed behind a horizon, as illustrated in Fig.~\ref{fig:cusp-collapse}.

We can demonstrate this explicitly by considering the behaviour of solutions to Eq.~\eqref{eq:EoM} near a cusp.
Choosing our coordinates such that the cusp occurs at $t=\sigma=0$, we can Taylor-expand the left- and right-moving modes as
    \begin{equation}
        \vb*X_\pm(\sigma_\pm)=\vu*n\sigma_\pm+\frac{1}{2}\ddot{\vb*X}_\pm\sigma_\pm^2+\frac{1}{6}\dddot{\vb*X}_\pm\sigma_\pm^3+\order{\sigma_\pm^4},
    \end{equation}
    where $\vu*n$ is a unit vector, and the higher derivatives are evaluated at the cusp.
(This is similar to the approach in Ref.~\cite{Damour:2001bk} for calculating the GW emission from cusps.)
Differentiating Eq.~\eqref{eq:left-right-constraint} gives the constraints
    \begin{equation}
    \label{eq:left-right-constraints-at-cusp}
        \vu*n\vdot\ddot{\vb*X}_\pm=0,\qquad\vu*n\vdot\dddot{\vb*X}_\pm=-|\ddot{\vb*X}_\pm|^2.
    \end{equation}
The position and velocity of the loop near the cusp at time $t=0$ are
    \begin{align}
    \begin{split}
    \label{eq:solution-near-cusp}
        \vb*X_0(\sigma)&\equiv\frac{1}{2}[\vb*X_+(\sigma)+\vb*X_-(-\sigma)]=\frac{1}{2}\ddot{\vb*X}\sigma^2+\order{\sigma^3},\\
        \dot{\vb*X}_0(\sigma)&\equiv\frac{1}{2}[\dot{\vb*X}_+(\sigma)+\dot{\vb*X}_-(-\sigma)]=\vu*n+\frac{1}{2}\dddot{\vb*X}\sigma^2+\order{\sigma^3},
    \end{split}
    \end{align}
    so that the distance from the cusp as a function of $\sigma$ is given by
    \begin{equation}
    \label{eq:distance-from-cusp}
        r_0(\sigma)=\sqrt{\vb*X_0\vdot\vb*X_0}=\frac{1}{2}|\ddot{\vb*X}|\sigma^2+\order{\sigma^3}.
    \end{equation}
We see that the fact that $\dot{\vb*X}_+=\dot{\vb*X}_-=\vu*n$ at the cusp means that there is no term of order $\sigma$ in Eq.~\eqref{eq:distance-from-cusp}, and the distance grows much more slowly for small $\sigma$ than it would on a non-cuspy part of the loop; this is the crucial ingredient for fulfilling the hoop condition.

Consider now a sphere of radius $r\ll\ell$.
We see from Eq.~\eqref{eq:distance-from-cusp} that the portion of the loop contained in the sphere is given by $-\sigma_*\le\sigma\le\sigma_*$, where $\sigma_*\ll\ell$ is defined by $r=r_0(\sigma_*)$, such that
    \begin{equation}
    \label{eq:sigma-star}
        \sigma_*=\qty(\frac{2r}{|\ddot{\vb*X}|})^{1/2}.
    \end{equation}
Using Eqs.~\eqref{eq:T00},~\eqref{eq:M-sphere}, and~\eqref{eq:sigma-star}, we see that the mass contained in the sphere is
    \begin{equation}
        M_\mathrm{sphere}=\mu\int_{-\sigma_*}^{+\sigma_*}\dd{\sigma}=2\mu\sigma_*=\qty(\frac{8\mu^2r}{|\ddot{\vb*X}|})^{1/2}.
    \end{equation}
The hoop condition~\eqref{eq:hoop-condition} is therefore satisfied if $r|\ddot{\vb*X}|\le32(G\mu)^2$, with the limiting PBH mass being
    \begin{equation}
    \label{eq:max-mass}
        M_\mathrm{PBH}=\frac{16G\mu^2}{|\ddot{\vb*X}|}.
    \end{equation}

The fact that Eq.~\eqref{eq:max-mass} depends on the cusp's acceleration $|\ddot{\vb*X}|$ rather than its velocity may seem surprising at first, but we can understand this intuitively by using Eqs.~\eqref{eq:EoM} and~\eqref{eq:tangent-vector-gamma} to write $|\ddot{\vb*X}|=|\vb*X''|=\dv{(1/\gamma)}{\sigma}$.
The acceleration therefore tells us about the rate of change of the Lorentz factor along the loop, and thus controls the size of the region that satisfies the hoop condition, which sets the PBH mass.
We can estimate the acceleration at the cusp by noting that generic solutions for $\vb*X$ can be written as a sum of Fourier modes,
    \begin{equation}
        \vb*X=\sum_{n=1}^\infty\qty[\vb*X_+^{(n)}\exp(\frac{2\uppi\mathrm{i}n\sigma_+}{\ell})+\vb*X_-^{(n)}\exp(\frac{2\uppi\mathrm{i}n\sigma_-}{\ell})].
    \end{equation}
Consider first the unrealistic case of a solution with a single mode $n$.
Since $|\dot{\vb*X}|=1$ at the cusp, we would then have $|\ddot{\vb*X}|=2\uppi n/\ell$.
For a more realistic solution, there are cross-terms from various different modes, but in general we can write $|\ddot{\vb*X}|=2\uppi\bar{n}/\ell$, where the ``effective mode number'' $\bar{n}$ is of order unity for smooth strings, and becomes larger for very wiggly strings.
One generally expects gravitational backreaction to dampen higher-order modes, which would dynamically drive $\bar{n}$ towards smaller values over time.

We therefore find that cusps lead to the formation of PBHs with mass
    \begin{equation}
    \label{eq:M_PBH}
        M_\mathrm{PBH}=\frac{8}{\uppi\bar{n}}G\mu^2\ell\approx G\mu M_\mathrm{loop},
    \end{equation}
    which are a factor of $G\mu$ smaller than those formed from circular collapse.
This is a generic prediction for \emph{all} Nambu-Goto loops.

\section{Properties of the PBHs}
\label{sec:pbh-properties}

\begin{figure}[t!]
    \includegraphics[width=0.48\textwidth]{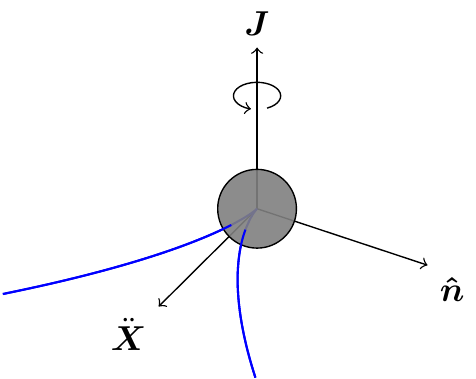}
    \caption{%
    An illustration of the PBH (in grey) immediately after formation.
    The cusp acceleration $\ddot{\vb*X}$, cusp velocity $\vu*n$, and PBH angular momentum $\vb*J$ are all orthogonal to each other.
    The cosmic string (in blue) punctures the horizon at two points separated by a small angle $\sim G\mu$, with its cusp hidden behind the horizon.}
    \label{fig:bh-cusp}
\end{figure}

We can estimate the properties of the PBHs formed through cusp collapse by assuming that all of the energy-momentum inside the sphere of radius $2GM$ at time $t=0$ is trapped behind the horizon.
Using Eq.~\eqref{eq:energy-momentum}, the PBH's linear and angular momenta are then given by
    \begin{align}
    \begin{split}
        P^i&=\int_{\mathcal{B}_r}\dd[3]{\vb*x}T^{0i}(0,\vb*x)=\mu\int_{-\sigma_*}^{+\sigma_*}\dd{\sigma}\dot{X}_0^i,\\
        J^i&=\int_{\mathcal{B}_r}\dd[3]{\vb*x}\varepsilon_{ijk}x^jT^{0k}(0,\vb*x)=\mu\int_{-\sigma_*}^{+\sigma_*}\dd{\sigma}\varepsilon_{ijk}X_0^j\dot{X}_0^k,
    \end{split}
    \end{align}
    where $\varepsilon_{ijk}$ is the 3-dimensional Levi-Civita symbol.
Inserting the leading-order terms from Eq.~\eqref{eq:solution-near-cusp}, and using $M=2\mu\sigma_*=16G\mu^2/|\ddot{\vb*X}|$, we find
    \begin{equation}
        \vb*P=M\qty[\vu*n+\frac{32(G\mu)^2}{3}\frac{\dddot{\vb*X}}{|\ddot{\vb*X}|^2}],\qquad \vb*J=\frac{2GM^2}{3}\frac{\ddot{\vb*X}}{|\ddot{\vb*X}|}\times\vu*n.
    \end{equation}
Thus we see that immediately after formation, the PBH is moving in the cusp direction $\vu*n$ with an ultrarelativistic velocity $v=|\vb*P|/M\approx1$.
In fact, the PBH's Lorentz factor is of the same order of magnitude as our estimate~\eqref{eq:gamma-condition},
    \begin{equation}
        \gamma\le\sqrt{\frac{3}{128}}(G\mu)^{-1},
    \end{equation}
    where we have used Eq.~\eqref{eq:left-right-constraints-at-cusp} and the triangle inequality $|\ddot{\vb*X}_+|^2+|\ddot{\vb*X}_-|^2\ge|\ddot{\vb*X}_++\ddot{\vb*X}_-|^2$.
We also see that the PBH is spinning around an axis orthogonal to both the cusp's velocity $\vu*n$ and its acceleration $\ddot{\vb*X}$, as illustrated in Fig.~\ref{fig:bh-cusp}, with a dimensionless spin parameter
    \begin{equation}
        \chi\equiv\frac{|\vb*J|}{GM^2}=2/3
    \end{equation}
    that is two-thirds of the extremal Kerr value $\chi=1$.
This large spin is a direct consequence of the orthogonality of the cusp's velocity and acceleration, as enforced by Eq.~\eqref{eq:left-right-constraints-at-cusp}.

Note that the mass $M$ includes the kinetic energy of the PBH, which is large due to its ultrarelativistic velocity.
The PBH's rest mass is given by
    \begin{equation}
    \label{eq:rest-mass}
        m\equiv\sqrt{M^2-P^2}=\frac{M}{\gamma}\approx(G\mu)^2\mu\ell,
    \end{equation}
    which is smaller than the total mass of the loop by a factor of $\sim(G\mu)^2$.
This means that the PBH radius is $\sim(G\mu)^3\ell$.
By using the Nambu-Goto approximation, we have assumed throughout that the cosmic strings have zero width, effacing any physics which occurs on lengthscales smaller than the string width $\delta\approx(\mu/\hbar)^{-1/2}$.
We therefore expect our results to hold only if the PBH radius is larger than $\delta$, which implies
    \begin{equation}
    \label{eq:ell-min-cusp-collapse}
        \ell\gtrsim\frac{\delta}{(G\mu)^3}=\frac{\ell_*}{(G\mu)^2}\approx5.1\,\mathrm{km}\,\times\qty(\frac{G\mu}{10^{-11}})^{-7/2},
    \end{equation}
    which corresponds to a minimum PBH rest mass of
    \begin{equation}
    \label{eq:M-min}
        m_\mathrm{min}\approx\frac{\delta}{G}\approx\eta\approx\frac{m_\mathrm{P}}{(G\mu)^{1/2}}\approx6.9\,\mathrm{g}\,\times\qty(\frac{G\mu}{10^{-11}})^{-1/2},
    \end{equation}
    where we recall that $\eta$ is the energy scale at which the cosmic strings are formed.
We thus see that there are three different classes of loops, corresponding to three broad ranges of loop lengths: loops smaller than $\ell_*\equiv\delta/(G\mu)$ are driven by particle radiation (as mentioned before), loops larger than $\ell_*/(G\mu)^2$ generically form PBHs from cusps, and loops inbetween are unchanged compared to the standard treatment in the literature.
These different regimes are summarised in Fig.~\ref{fig:scales}.

The loop punctures the PBH horizon at two points, corresponding to the position of the loop at $(t,\sigma)=(0,\pm\sigma_*)$.
Using the expansion~\eqref{eq:solution-near-cusp} for $\vb*X_0(\pm\sigma_*)$, we see that both points lie very near to the $\ddot{\vb*X}$ axis, which is in the PBH's equatorial plane.
By continuing the expansion to at least $\order{\sigma^3}$, one can show that
    \begin{equation}
        \cos\theta_*\equiv\frac{\vb*X_0(\sigma_*)\vdot\vb*X_0(-\sigma_*)}{|\vb*X_0(\sigma_*)||\vb*X_0(-\sigma_*)|}=1+\order{\sigma_*^2},
    \end{equation}
    which implies that the angle between the two puncture points is $\theta_*\sim\sigma_*/\ell\sim G\mu$, and thus that both points are very close to the equatorial plane.
(This is important for our discussion of the subsequent dynamics of the loop near the PBH in Section~\ref{sec:loop-pbh-dynamics}.)

Accounting for the finite string width $\delta$, we see that these two puncture points are so close that it is possible for the loop to self-intersect at the PBH horizon.
Using simple trigonometry, the separation between the puncture points on the horizon is roughly
    \begin{equation}
    \label{eq:puncture-separation}
        2Gm_\mathrm{PBH}\tan\frac{\theta_*}{2}\sim(G\mu)^4\ell,
    \end{equation}
    so the loop self-intersects at the horizon if this separation is smaller than the string width $\delta$, which occurs if
    \begin{equation}
    \label{eq:puncture-intersection}
        \ell\lesssim\frac{\delta}{(G\mu)^4}=\frac{\ell_*}{(G\mu)^3}\approx1.7\times10^{-2}\,\mathrm{pc}\times\qty(\frac{G\mu}{10^{-11}})^{-9/2}.
    \end{equation}
In this case, one would expect the string to intercommute near the horizon, meaning that the PBH would be immediately chopped off from the loop at formation.

\section{PBHs from pseudocusp collapse}
\label{sec:pseudocusp-collapse}

Consider now a generic loop segment at some time $t=0$, whose configuration is locally described by
    \begin{equation}
    \label{eq:loop-configuration}
        \vb*X_\pm(\sigma_\pm)=\vu*n_\pm\sigma_\pm+\frac{1}{2}\ddot{\vb*X}_\pm\sigma_\pm^2+\order{\sigma_\pm^3}.
    \end{equation}
If $\vu*n_+=\vu*n_-$, the point $\sigma=0$ is a cusp with a divergent Lorentz factor $\gamma\to\infty$, which forms a PBH as described above.
Our simple argument in Eq.~\eqref{eq:gamma-condition} suggests that a divergent Lorentz factor is sufficient but not necessary for PBH formation; all we need is $\gamma=\order{1/G\mu}$.
We therefore expect PBH formation in situations where $\kappa\equiv|\vu*n_+-\vu*n_-|$ is nonzero, so long as $\kappa$ is small enough.
We refer to points on the loop where $\kappa$ is small but nonzero as ``pseudocusps.''

In order to estimate how small $\kappa$ must be, we generalise Eq.~\eqref{eq:distance-from-cusp} to give the distance from the pseudocusp for small $\sigma$ at time $t=0$,
    \begin{equation}
        r_0^2(\sigma)\simeq\frac{\sigma^2}{2}(1-\vu*n_+\vdot\vu*n_-)+\frac{\sigma^3}{4}(\vu*n_+\vdot\ddot{\vb*X}_--\vu*n_-\vdot\ddot{\vb*X}_+)+\frac{\sigma^4}{4}|\ddot{\vb*X}|^2.
    \end{equation}
Defining the angle between the left- and right-moving mode velocities, $\cos^{-1}\vu*n_+\vdot\vu*n_-=\kappa+\order{\kappa^3}$, we find to leading order that
    \begin{equation}
    \label{eq:distance-from-pseudocusp}
        r_0(\sigma)\approx\frac{1}{2}\kappa|\sigma|+\frac{1}{2}|\ddot{\vb*X}|\sigma^2,
    \end{equation}
    where we have used the constraints~\eqref{eq:left-right-constraints-at-cusp}, and have approximated $(\vu*n_+-\vu*n_-)\vdot\ddot{\vb*X}\approx\kappa|\ddot{\vb*X}|$.
Repeating the arguments leading to Eq.~\eqref{eq:max-mass}, we find that a PBH forms so long as $\kappa<8G\mu$, with the corresponding mass given by
    \begin{equation}
        M=\frac{2\mu}{|\ddot{\vb*X}|}(8G\mu-\kappa).
    \end{equation}
The loop velocity at $\sigma=0$ is $v\simeq1-\kappa^2/8$, so we can translate the bound $\kappa<8G\mu$ into a bound on the pseudocusp velocity.
Doing so, we see that pseudocusps collapse to form PBHs so long as their Lorentz factor obeys
    \begin{equation}
    \label{eq:gamma-condition-pseudocusp}
        \gamma\ge\frac{1}{4G\mu},
    \end{equation}
    in agreement with our simple estimate~\eqref{eq:gamma-condition}.
Since pseudocusps occur even more generically on loops than cusps do~\cite{Stott:2016loe}, this result further enhances the PBH formation rate.

Note that in writing the Taylor expansion~\eqref{eq:loop-configuration} we have assumed that the loop is smooth in the neighbourhood of $\sigma=0$, which precludes any discontinuities in the loop's tangent vector (commonly called \emph{kinks}).
However, it is easy to convince oneself that kinks do not contribute to PBH formation.
A kink at some $\sigma_\mathrm{k}$ near $\sigma=0$ would make Eq.~\eqref{eq:distance-from-pseudocusp} a piecewise smooth function, with different coefficients for each order in $\sigma$ on either side of the kink.
Generically these coefficients are of the same order of magnitude on both sides of the kink; there is nothing about the kink which forces the $\order{\sigma}$ term in Eq.~\eqref{eq:distance-from-pseudocusp} to be small, which is what we require for PBH formation.
As we have shown above, the smallness of this term is uniquely associated with a large Lorentz factor, and therefore with (pseudo)cusps.

Of course, this last argument depends strongly on the Nambu-Goto approximation; in a full field-theoretic setting one would expect kinks to carry gradient energy, which may be sufficiently concentrated to satisfy the hoop condition.
However, one would only expect the gradient energy to be large in a region of size comparable to the string width $\delta$, meaning the resulting PBH masses would be near the minimal mass $m_\mathrm{min}\sim\delta/G$ from Eq.~\eqref{eq:M-min}.
For kinks, as for cusps, we can trust the Nambu-Goto approximation so long as we consider PBHs with mass $m\gg m_\mathrm{min}$.

\section{Backreaction on the cusp}
\label{sec:backreaction}

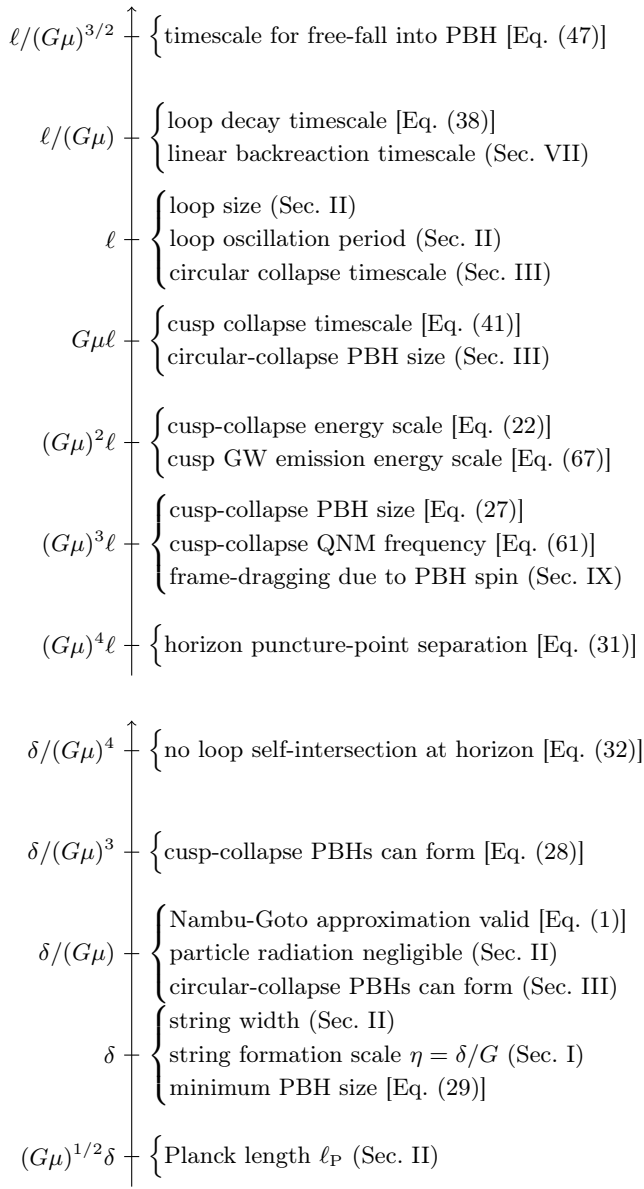
\begin{figure}[t!]
    \begin{tikzpicture}[scale=10pt]
        \draw[->] (0,0.025) -- (0,0.915);
        \draw[-]  (0.01,0.875) -- (-0.01,0.875) node[anchor=east]{$\ell/(G\mu)^{3/2}$};
        \node[right] at (0.01,0.875) {$
            \begin{cases}
                \text{timescale for free-fall into PBH [Eq.~\eqref{eq:free-fall-time}]}
            \end{cases}$};
        \draw[-]  (0.01,0.74) -- (-0.01,0.74) node[anchor=east]{$\ell/(G\mu)$};
        \node[right] at (0.01,0.74) {$
            \begin{cases}
                \text{loop decay timescale [Eq.~\eqref{eq:decay-time}]}\\
                \text{linear backreaction timescale (Sec.~\ref{sec:backreaction})}
            \end{cases}$};
        \draw[-]  (0.01,0.605) -- (-0.01,0.605) node[anchor=east]{$\ell$};
        \node[right] at (0.01,0.605) {$
            \begin{cases}
                \text{loop size (Sec.~\ref{sec:nambu-goto})}\\
                \text{loop oscillation period (Sec.~\ref{sec:nambu-goto})}\\
                \text{circular collapse timescale (Sec.~\ref{sec:hoop-conjecture})}
            \end{cases}$};
        \draw[-]  (0.01,0.47) -- (-0.01,0.47) node[anchor=east]{$G\mu\ell$};
        \node[right] at (0.01,0.47) {$
            \begin{cases}
                \text{cusp collapse timescale [Eq.~\eqref{eq:collapse-timescale}]}\\
                \text{circular-collapse PBH size (Sec.~\ref{sec:hoop-conjecture})}
            \end{cases}$};
        \draw[-]  (0.01,0.335) -- (-0.01,0.335) node[anchor=east]{$(G\mu)^2\ell$};
        \node[right] at (0.01,0.335) {$
            \begin{cases}
                \text{cusp-collapse energy scale [Eq.~\eqref{eq:M_PBH}]}\\
                \text{cusp GW emission energy scale [Eq.~\eqref{eq:gw-radiation-fraction}]}
            \end{cases}$};
        \draw[-]  (0.01,0.2) -- (-0.01,0.2) node[anchor=east]{$(G\mu)^3\ell$};
        \node[right] at (0.01,0.2) {$
            \begin{cases}
                \text{cusp-collapse PBH size [Eq.~\eqref{eq:rest-mass}]}\\
                \text{cusp-collapse QNM frequency [Eq.~\eqref{eq:qnm-omega}]}\\
                \text{frame-dragging due to PBH spin (Sec.~\ref{sec:loop-pbh-dynamics})}
            \end{cases}$};
        \draw[-]  (0.01,0.065) -- (-0.01,0.065) node[anchor=east]{$(G\mu)^4\ell$};
        \node[right] at (0.01,0.065) {$
            \begin{cases}
                \text{horizon puncture-point separation [Eq.~\eqref{eq:puncture-separation}]}
            \end{cases}$};
        \draw[->] (0,-0.655) -- (0,-0.035);
        \draw[-]  (0.01,-0.075) -- (-0.01,-0.075) node[anchor=east]{$\delta/(G\mu)^4$};
        \node[right] at (0.01,-0.075) {$
            \begin{cases}
                \text{no loop self-intersection at horizon [Eq.~\eqref{eq:puncture-intersection}]}
            \end{cases}$};
        \draw[-]  (0.01,-0.21) -- (-0.01,-0.21) node[anchor=east]{$\delta/(G\mu)^3$};
        \node[right] at (0.01,-0.21) {$
            \begin{cases}
                \text{cusp-collapse PBHs can form [Eq.~\eqref{eq:ell-min-cusp-collapse}]}
            \end{cases}$};
        \draw[-]  (0.01,-0.345) -- (-0.01,-0.345) node[anchor=east]{$\delta/(G\mu)$};
        \node[right] at (0.01,-0.345) {$
            \begin{cases}
                \text{Nambu-Goto approximation valid [Eq.~\eqref{eq:ell-star}]}\\
                \text{particle radiation negligible (Sec.~\ref{sec:nambu-goto})}\\
                \text{circular-collapse PBHs can form (Sec.~\ref{sec:hoop-conjecture})}
            \end{cases}$};
        \draw[-]  (0.01,-0.48) -- (-0.01,-0.48) node[anchor=east]{$\delta$};
        \node[right] at (0.01,-0.48) {$
            \begin{cases}
                \text{string width (Sec.~\ref{sec:nambu-goto})}\\
                \text{string formation scale }\eta=\delta/G\text{ (Sec.~\ref{sec:intro})}\\
                \text{minimum PBH size [Eq.~\eqref{eq:M-min}]}
            \end{cases}$};
        \draw[-]  (0.01,-0.615) -- (-0.01,-0.615) node[anchor=east]{$(G\mu)^{1/2}\delta$};
        \node[right] at (0.01,-0.615) {$
            \begin{cases}
                \text{Planck length }\ell_\mathrm{P}\text{ (Sec.~\ref{sec:nambu-goto})}
            \end{cases}$};
    \end{tikzpicture}
    \caption{%
    A summary of the different scales in the cusp-collapse problem, including references to the equation or section where they first appear.
    In the Nambu-Goto approximation there are only two dimensionful quantities: the string tension $\mu$ and the loop length $\ell$.
    Since the string tension usually appears in the dimensionless combination $G\mu$, all of the system's time- and length-scales can be written as $(G\mu)^p\times\ell$ for some power $p$.
    The fact that $G\mu\ll1$ means that there is a strong hierarchy between these scales.
    Going beyond Nambu-Goto introduces another dimensionful quantity with its own hierarchy of scales: the string width $\delta\approx(\mu/\hbar)^{-1/2}$.
    (Note that many of the scales associated with $\delta$ here are lower limits on the loop size $\ell$; e.g., $\delta/(G\mu)$ is the smallest loop size for which the Nambu-Goto approximation is valid.)
    These two sets of scales are shifted relative to each other depending on $\ell$.
    More complicated combinations of $\ell$ and $\delta$ are of course possible; e.g., the evaporation timescales for cusp-collapse and circular-collapse PBHs are $(G\mu)^8\ell^3/\delta^2$ and $(G\mu)^2\ell^3/\delta^2$ respectively.}
    \label{fig:scales}
\end{figure}

One of the main assumptions of our analysis is that the loop is described by the flat-space equations of motion~\eqref{eq:EoM} right up to the instant of PBH formation.
A more complete analysis would account for the gravitational backreaction of the loop on its own dynamics, which one would expect to suppress the cusp.
One might worry whether this suppression is strong enough to prevent the PBH from forming.

Significant evidence against this worry comes from the extensive literature on cosmic-string backreaction~\cite{Thompson:1988yj,Quashnock:1990wv,Copeland:1990qu,Battye:1994qa,Buonanno:1998is,Carter:1998ix,Wachter:2016hgi,Wachter:2016rwc,Blanco-Pillado:2018ael,Chernoff:2018evo,Blanco-Pillado:2019nto}, in which numerous different approaches (both analytical and numerical) have repeatedly shown that backreaction does not prevent cusps from forming.
There is general agreement that cusps are suppressed to some degree by backreaction, but that this suppression occurs gradually over many loop oscillation periods, on a timescale of order the loop decay time
    \begin{equation}
    \label{eq:decay-time}
        t_\mathrm{decay}\sim\frac{\ell}{G\mu}.
    \end{equation}

A serious problem with this argument is that essentially all of the existing work on string backreaction has been done in the weak-field limit, treating the string's gravity as a linear perturbation on the background spacetime.
This linearised approach is clearly unable to capture strong-gravity effects such as PBH formation, which explains why cusp collapse has not been identified previously.

We can reassure ourselves by considering the timescale on which the PBH formation occurs.
The velocity of the string point $\sigma=0$ at times near to the cusp, $|t|\ll\ell$, can be written as
    \begin{equation}
        \dot{\vb*X}_\mathrm{c}(t)=\vu*n+t\ddot{\vb*X}+\frac{1}{2}t^2\dddot{\vb*X}+\order{t^3},
    \end{equation}
    with the corresponding Lorentz factor given by
    \begin{equation}
    \label{eq:gamma-cusp}
        \gamma_\mathrm{c}(t)\simeq\frac{2}{|t|}|\ddot{\vb*X}_+-\ddot{\vb*X}_-|^{-1}\approx\frac{\ell}{\uppi\bar{n}|t|},
    \end{equation}
    where we have used the constraints~\eqref{eq:left-right-constraints-at-cusp}, and the last equality generally holds to within an order of magnitude.
Since PBH formation is associated with the Lorentz factor growing above a certain threshold~\eqref{eq:gamma-condition-pseudocusp}, we can estimate the associated timescale by setting Eq.~\eqref{eq:gamma-cusp} equal to this threshold, giving
    \begin{equation}
    \label{eq:collapse-timescale}
        \upDelta t_\mathrm{PBH}\sim G\mu\ell.
    \end{equation}
This shows that the PBH is formed on an extremely short timescale: shorter than the loop oscillation period by a factor of $G\mu$, and shorter than the timescale for linear backreaction by a factor of $(G\mu)^2$.
(See Fig.~\ref{fig:scales} for an overview of the different time- and length-scales in the loop-PBH system.)
Even if nonlinear backreaction is in principle strong enough to prevent the cusp from forming, it seems unlikely that it can act on a short enough timescale to do so, meaning that backreaction is unlikely to prevent PBH formation.\footnote{%
See also Ref.~\cite{Thompson:1988yj}, which examines backreaction on cusps without resorting to linearised gravity, and argues geometrically that cusps form \emph{``no matter how strong the gravitational field near a cusp''}.}

\section{Radiation from the collapse}
\label{sec:radiation}

Our analysis thus far has neglected the effects of gravitational radiation during the collapse.
Radiation is likely to be important, as the collapse is ultrarelativistic and highly nonspherical.
In general, one would expect the final mass, linear momentum, and angular momentum of the PBH to be of the form
    \begin{equation}
    \label{eq:radiation-efficiency}
        M=M_0(1-\epsilon_M),\quad P=P_0(1-\epsilon_P),\quad J=J_0(1-\epsilon_J),
    \end{equation}
    where a zero subscript denotes the na\"{i}ve, zero-radiation quantity calculated above, and the $\epsilon_i$ are three numbers between zero and unity, describing the efficiency with which each quantity is radiated away.
In the context of circular loop collapse, Hawking~\cite{Hawking:1990tx} calculated a theoretical upper bound on $\epsilon_M$ of $1-\sqrt{1/2}\approx29\%$ by explicitly constructing a marginally outer trapped surface in the spacetime of the collapsing loop and requiring that this surface be enclosed by the event horizon of the final PBH.
This argument depends heavily on the circular symmetry of the loop, and no such construction seems possible in our case.

Despite the lack of symmetries here, we can make some interesting statements by requiring that the final PBH spin $\chi=J/(GM^2)$ be less than or equal to unity; otherwise the PBH would be ``overspun'' to reveal a naked singularity, violating cosmic censorship~\cite{Penrose:1973um,Wald:1997wa}.
Since $\chi_0=2/3$, we can write
    \begin{equation}
    \label{eq:spin-bound}
        \chi=\frac{2}{3}\frac{(1-\epsilon_J)}{(1-\epsilon_M)^2}\le1.
    \end{equation}
We see that, so long as $\epsilon_J\lesssim2\epsilon_M$, the final spin parameter of the PBH is larger than the na\"{i}ve value $2/3$, which shows that the upper bound~\eqref{eq:spin-bound} is likely to be useful.
In general, we expect $\epsilon_J\lesssim\epsilon_M$; see e.g. Ref.~\cite{Durrer:1989zi}, in which the rate at which loops radiate angular momentum is shown to be typically an order of magnitude smaller than the rate at which they radiate mass.
If $\epsilon_J=\epsilon_M$, then Eq.~\eqref{eq:spin-bound} gives
    \begin{equation}
    \label{eq:spin-bound-33}
        \epsilon_M\le1/3\approx33\%.
    \end{equation}
In the limit where $\epsilon_J\to0$, the bound is even stronger,
    \begin{equation}
    \label{eq:spin-bound-18}
        \epsilon_M\le1-\sqrt{2/3}\approx18\%.
    \end{equation}
Since we expect $0<\epsilon_J<\epsilon_M$, the true upper bound for cusp collapse is likely to lie somewhere between Eqs.~\eqref{eq:spin-bound-33} and~\eqref{eq:spin-bound-18}.

Interestingly, numerical relativity simulations of circular loop collapse performed in Refs.~\cite{Helfer:2018qgv,Aurrekoetxea:2020tuw} found $\epsilon_M\lesssim2\%$, well below Hawking's bound.
The authors suggest that this is due to the symmetry of the circular collapse, which means the horizon is nearly spherical when it first forms, suppressing the total radiation.
The initial horizon in our case is likely to be highly distorted, meaning that $\epsilon_M$ is likely to be closer to its upper bound.
Of course, it is possible that cusp collapse radiates angular momentum much more efficiently than is typical for loops as a whole, in which case $\epsilon_J$ could be larger than in Ref.~\cite{Durrer:1989zi}.
It would then be possible for $\epsilon_M$ to be larger than the rough bounds in Eqs.~\eqref{eq:spin-bound-33} and~\eqref{eq:spin-bound-18}.

We note in passing that radiation of linear momentum ($\epsilon_P>0$) would lead to a ``rocket effect''~\cite{Hogan:1984is,Vachaspati:1984gt}, in which the loop's centre of mass is given a kick in the opposite direction to the radiation.
However, even if this process is maximally efficient, the radiated momentum is at most $P_0\approx G\mu M_\mathrm{loop}$, so the maximum kick is $v\approx G\mu$.
This pales in comparison to the rms velocity of points on the loop, $v_\mathrm{rms}=\sqrt{1/2}\approx0.707$, so the effect is of negligible interest; radiation from elsewhere on the loop quickly cancels out the kick.

\section{Dynamics of the loop-PBH system}
\label{sec:loop-pbh-dynamics}

Once formed, the PBH is inextricably linked to the surrounding string; the portion of the string enclosed behind the horizon cannot escape, and the portion outside the horizon is topologically forbidden from detaching itself.
Since the mass of our PBHs is smaller than that of their parent loops by a factor of $G\mu$, we expect the loop dynamics to be largely unaffected by the presence of the PBH, at least on timescales $\sim\ell$.
In particular, this means that despite its ultrarelativistic velocity, the PBH cannot drag the rest of the loop along with it---instead, we expect the loop's tension to act on the PBH to decelerate it, and for the loop to continue oscillating in essentially the same motion as before.
This could mean that cusp-collapse PBHs do not trace the DM distribution, as their parent loops could easily drag them out of DM haloes.
(This possibility was also pointed out in Ref.~\cite{Vilenkin:2018zol}, albeit for a different PBH formation scenario.)

Most cusp-collapse PBHs are very small, and decay rapidly through Hawking radiation~\cite{Hawking:1974sw}.
In particular, the evaporation timescale for the minimum mass~\eqref{eq:M-min} is $\approx t_\mathrm{P}(m_\mathrm{PBH}/m_\mathrm{P})^3\approx10^{-27}\,\mathrm{s}\times(G\mu/10^{-11})^{-3/2}$~\cite{Hawking:1974rv,Page:1976df,Page:1976ki}.
It is unclear what effect the loop has on the evaporation process, and vice versa.
The PBH cannot maintain its mass at the minimum value in Eq.~\eqref{eq:M-min} by accreting the loop, since this would correspond to the loop losing mass at a rate
    \begin{equation}
        \dv{m}{t}\approx\qty(\frac{m_\mathrm{P}}{m_\mathrm{PBH}})^2\frac{m_\mathrm{P}}{t_\mathrm{P}}\approx\mu\approx\frac{M_\mathrm{loop}}{\ell},
    \end{equation}
    i.e. the loop would have to lose all of its mass within a single oscillation period.
This seems very unlikely, given the limited gravitational influence of the PBH---the timescale for an object to free-fall from a distance $\sim\ell$ into the PBH is
    \begin{equation}
    \label{eq:free-fall-time}
        t_\mathrm{ff}\sim\frac{\ell^{3/2}}{\sqrt{Gm_\mathrm{PBH}}}\approx\frac{\ell}{(G\mu)^{3/2}},
    \end{equation}
    so even neglecting the loop's kinetic energy, it would take many oscillation periods for it to be accreted.
We are therefore forced to allow the PBH to decay to sizes smaller than the loop width.
It is hard to envisage a way for the topologically-stable field configuration around the string to be disrupted by the PBH evaporation, so the most likely outcome seems to be that the PBH simply vanishes from the loop.\footnote{%
Refs.~\cite{Bonjour:1998rf,Gregory:2013xca} found that abelian-Higgs string-BH systems can exhibit interesting ``flux expulsion'' effects when the BH is smaller than the string width; however, these results are only valid for extremal Kerr and Reissner-Nordstr{\"o}m BHs, and it is not clear whether they have any bearing on our sub-extremal PBHs, or on the evaporation process.}

PBHs with rest mass larger than $m_*\approx5\times10^{14}\,\mathrm{g}\approx3\times10^{-19}M_\odot$ evaporate very slowly, and lose a negligible fraction of their mass within a Hubble time~\cite{Carr:2020gox}.
It is therefore interesting to consider how these non-evaporating PBHs interact with their parent loops on cosmological timescales.
For the simplest case of an infinitely long straight string, explicit solutions for the metric of a BH threaded by a cosmic string have been constructed for the Nambu-Goto case~\cite{Aryal:1986sz} and for abelian-Higgs strings~\cite{Achucarro:1995nu,Chamblin:1997gk,Bonjour:1998rf,Dehghani:2001nz,Ghezelbash:2001pq,Gregory:2013xca}.
For solutions where the BH is rotating, the string is assumed to be aligned with the spin axis.
In each case the solution is static, and the string represents a form of stable long-range hair on the BH.
Since the string is static, its only gravitational effect is to induce a conical singularity along its axis, with a deficit angle $\sim G\mu$~\cite{Vilenkin:1981zs,Garfinkle:1985hr}.
This deficit angle means that the string-BH solution is not asymptotically flat, which explains how it evades the no-hair theorem~\cite{Ruffini:1971bza,Bekenstein:1996pn}.
The deficit angle can also modify the BH's quasi-normal mode spectrum~\cite{Chen:2008zzv,Cheung:2020dxo}.

The relevance of these results is somewhat limited in our case, as the string emanating from the PBH is not static, but continues to oscillate relativistically.
Perhaps even more importantly, the string is not locally aligned with the PBH's spin axis, so does not puncture the PBH at its poles like in Refs.~\cite{Aryal:1986sz,Achucarro:1995nu,Chamblin:1997gk,Bonjour:1998rf,Dehghani:2001nz,Ghezelbash:2001pq,Gregory:2013xca}.
Instead, due to the geometry of the cusp, the two points where the string punctures the horizon lie in---or very close to---the equatorial plane, and are separated by a small angle $\sim G\mu$ (as we showed in Sec.~\ref{sec:pbh-properties}).
Being in the equatorial plane, one would expect relativistic frame-dragging to pull the string into a spiral configuration around the PBH spin axis on scales $\sim Gm_\mathrm{PBH}$.
(On larger scales $\sim\ell$, the string tension easily overcomes the frame-dragging forces.)
This spiralling of the string around the PBH, combined with the very small separation between the two points at which it punctures the horizon, makes it seem likely that the string intersects itself near to the PBH.
The PBH would thus be chopped off from the rest of the loop, leaving it with only a small segment of string still attached, which it would rapidly accrete.
For sufficiently small loops, we have shown in Eq.~\eqref{eq:puncture-intersection} that the PBH is likely to be immediately chopped off at the moment of formation.

It is interesting to ask whether two PBHs connected to the same loop could have a greater chance of merging due to the loop dynamics; a similar effect has been demonstrated for the annihilation of monopole-antimonopole pairs connected by strings (so-called ``cosmic necklaces'')~\cite{Berezinsky:1997td,Siemens:2000ty}.
However, the two PBHs would likely be separated by a distance $\sim\ell$ much larger than their size, so based on the discussion above we would expect the PBHs to be chopped off before the loop has the chance to pull them together.

There are clearly many uncertainties in how cusp-collapse PBHs affect the loop network, but our very rough arguments here suggest that small PBHs rapidly evaporate to leave the loop essentially unchanged (although its dynamics are affected by the radiation), while large PBHs are likely to be cut off from loop by string self-intersections on small lengthscales $\sim Gm_\mathrm{PBH}\sim(G\mu)^3\ell$.

\section{The PBH mass spectrum}
\label{sec:mass-spectrum}

\begin{figure}[t!]
    \includegraphics[width=0.48\textwidth]{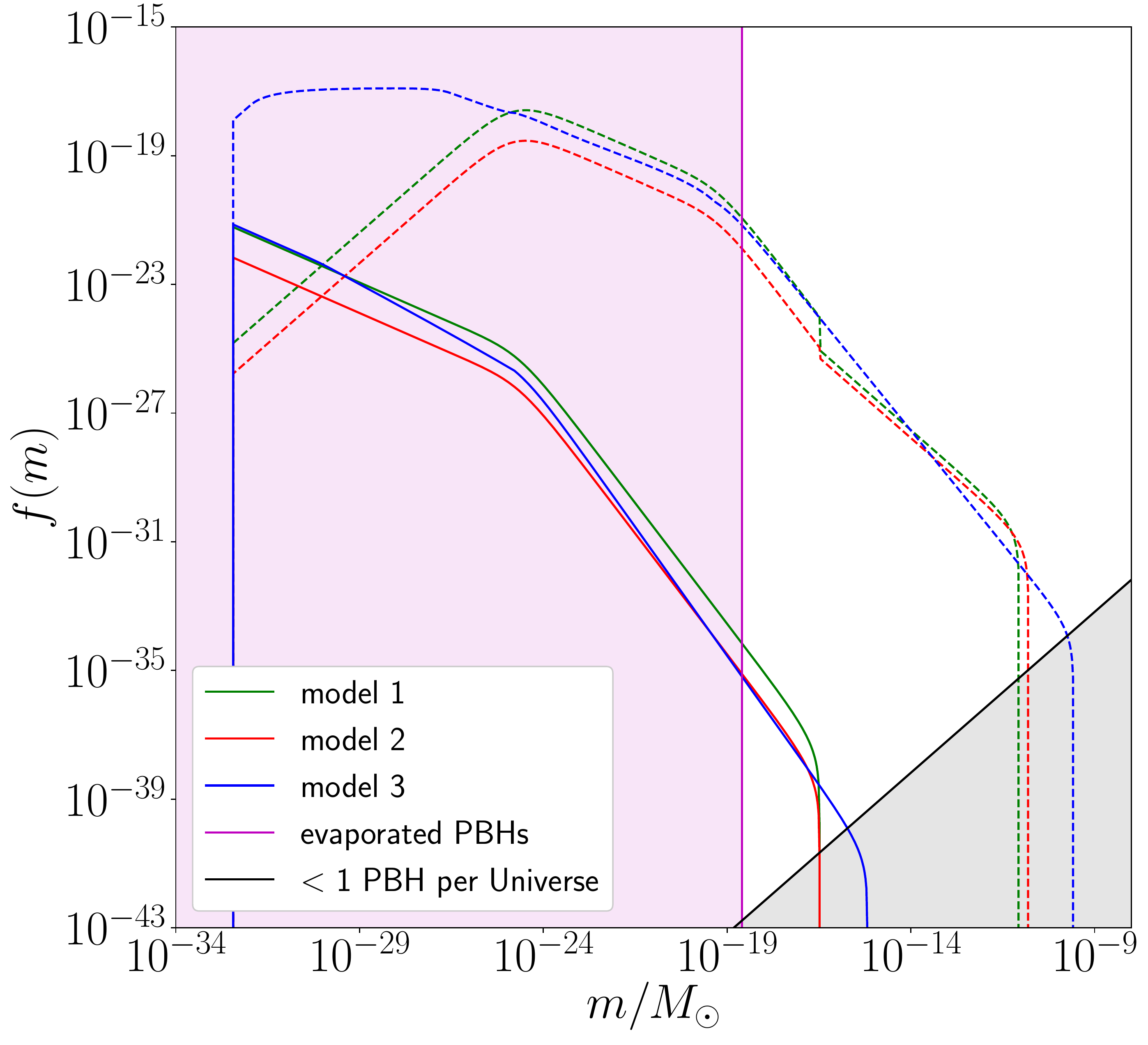}
    \caption{%
    The present-day PBH mass spectrum \eqref{eq:mass-spectrum} for three models of the cosmic string loop network with $G\mu=10^{-11}$.
    The solid and dashed lines correspond to PBHs formed in the radiation and matter eras respectively.
    The cutoffs at small and large masses are given by Eqs.~\eqref{eq:M-min} and~\eqref{eq:M_max}.
    The magenta region represents PBHs which have evaporated by the present day.
    The grey region corresponds to there being less than one PBH of that mass in the observable Universe.}
    \label{fig:mass-spectra-Gmu-1e-11}
\end{figure}

Cosmic string loops typically form one cusp per oscillation period, which means that cusp-collapse PBHs are continuously created by the loop network, from the very early Universe to the present day, resulting in a broad distribution of PBH masses.
This contrasts sharply with the standard PBH formation scenario, where the collapse typically occurs at a single early epoch, resulting in a monochromatic PBH mass spectrum.

We can write the comoving number density of cusp-collapse PBHs with rest mass between $m$ and $m+\dd{m}$ as
    \begin{equation}
        n_\mathrm{PBH}(m,t)\dd{m}=\int_0^t\dd{t'}\frac{2N_\mathrm{cusp}}{\ell_m}n_\mathrm{loop}(\ell_m,t')\dd{\ell_m},
    \end{equation}
    where $t$ is cosmic time, $N_\mathrm{cusp}$ is the average number of cusps per loop oscillation period, $n_\mathrm{loop}(\ell,t)\dd{\ell}$ is the comoving number density of loops with length between $\ell$ and $\ell+\dd{\ell}$, and $\ell_m\approx Gm/(G\mu)^3$ is the loop length required to form a PBH of mass $m$.
We assume $N_\mathrm{cusp}=1$, consistent with much of the literature on cosmic string phenomenology, particularly regarding GW searches~\cite{Abbott:2017mem,Auclair:2019wcv}.
The factor of $2/\ell_m$ here accounts for the oscillation period of the loop which forms the PBH.
Since $\ell_m$ corresponds to a fixed physical (rather than comoving) scale, PBH production only begins once this scale has entered the horizon.
This happens at cosmic time
    \begin{equation}
    \label{eq:collapse-time}
        t_i(m)\approx\ell_m\approx16\,\mathrm{Gyr}\times\frac{m}{10^{-10}M_\odot}\times\qty(\frac{G\mu}{10^{-11}})^{-3},
    \end{equation}
    which means that the largest PBHs form at the present day, with mass
    \begin{equation}
    \label{eq:M_max}
        m_\mathrm{max}\approx0.88\times10^{-10}\,M_\odot\times\qty(\frac{G\mu}{10^{-11}})^3.
    \end{equation}
Coincidentally, this corresponds to the ``sublunar'' mass range---one of the few regimes where there are no constraints preventing PBHs from making up the totality of DM~\cite{Carr:2020gox,Carr:2020xqk}.
However, the majority of cusp-collapse PBHs have masses much smaller than~\eqref{eq:M_max}.

The loop distribution usually evolves toward a scaling solution~\cite{Bennett:1987vf},
    \begin{equation}
        n_\mathrm{loop}(\ell,t)=\frac{a^3(t)}{t^4}\mathcal{F}(\gamma),
    \end{equation}
    where the distribution function $\mathcal{F}$ depends only on the dimensionless length $\gamma\equiv\ell/t$.
There are three widely-used models for the distribution function: the one-scale model of Refs.~\cite{Vilenkin:2000jqa,Siemens:2006vk}; the model of Ref.~\cite{Blanco-Pillado:2013qja}, which is calibrated to numerical simulations; and the model of Refs.~\cite{Ringeval:2005kr,Lorenz:2010sm}, which is calibrated to a different set of simulations, and includes additional modelling of the effects of backreaction on the loops.
Following Refs.~\cite{Abbott:2017mem,Auclair:2019wcv}, we refer to these as ``model 1,'' ``model 2,'' and ``model 3'' respectively.
Model 1 is widely considered obsolete, as it is incompatible with both of the main sets of Nambu-Goto network simulations~\cite{Ringeval:2005kr,Blanco-Pillado:2013qja}; however, we include it here for completeness.

The present-day PBH abundance is conveniently expressed in terms of
    \begin{equation}
        f(m)\equiv\frac{1}{\rho_\mathrm{CDM}}\dv{\rho_\mathrm{PBH}}{(\ln m)}=\frac{m^2n_\mathrm{PBH}(m)}{\rho_\mathrm{CDM}},
    \end{equation}
    which gives the mass density of PBHs in a logarithmic mass interval as a fraction of the total mass density of cold DM, $\rho_\mathrm{CDM}$.
For cusp collapse, this fraction is given by
    \begin{equation}
    \label{eq:mass-spectrum}
        f(m)=\frac{2N_\mathrm{cusp}m}{\rho_\mathrm{CDM}}\int_0^{t_0}\dd{t}\frac{a^3(t)}{t^4}\mathcal{F}\qty(\frac{Gm}{(G\mu)^3t}).
    \end{equation}
In general the integral in Eq.~\eqref{eq:mass-spectrum} is broken into two parts, corresponding to the different scaling solutions in the matter and radiation eras.

\begin{figure}[t!]
    \includegraphics[width=0.48\textwidth]{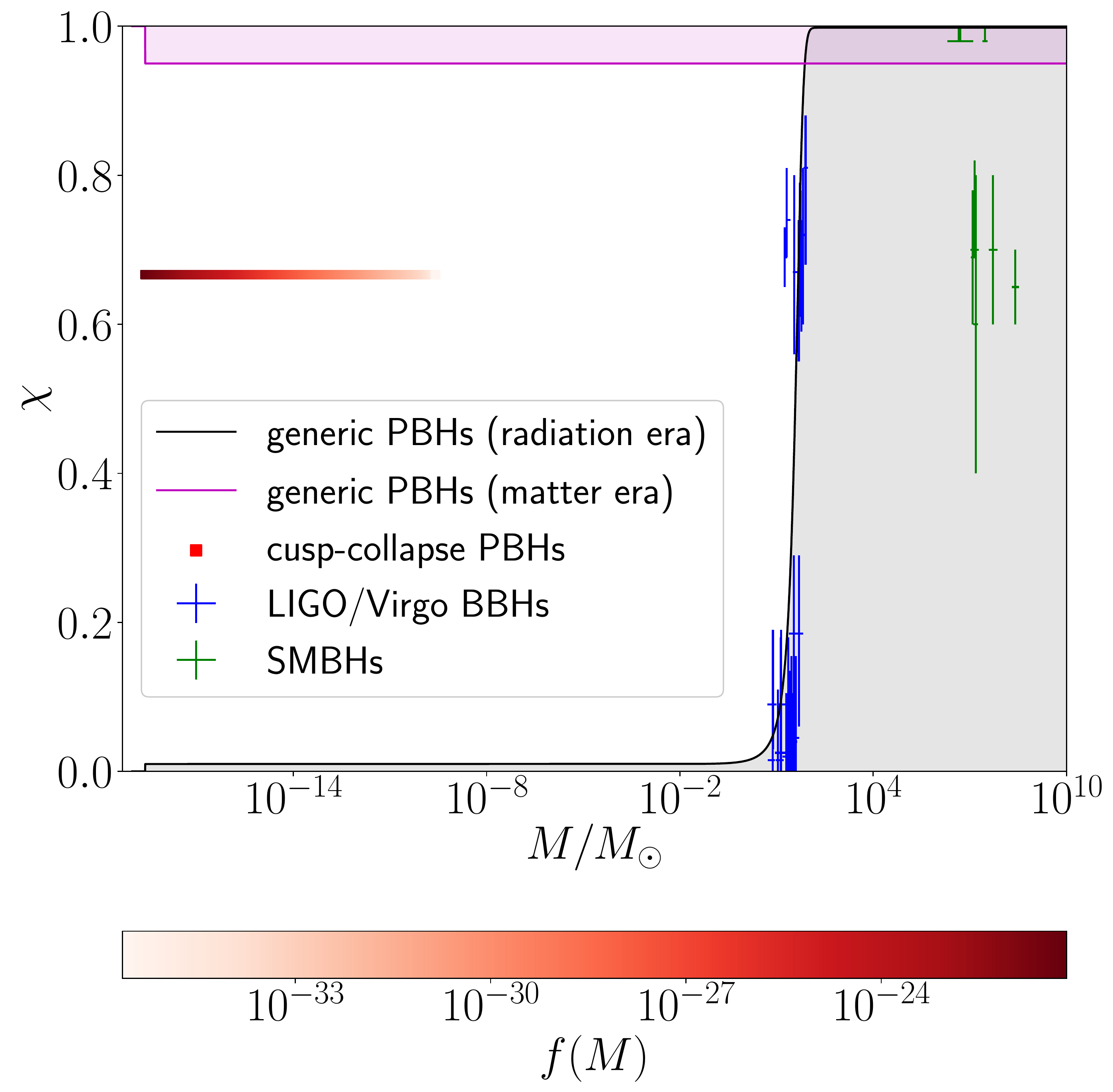}
    \caption{%
    The location of various primordial and astrophysical BH populations in the Regge plane (BH mass-spin parameter space).
    Blue crosses show the initial and final BHs for each of the ten binary BH mergers in LIGO/Virgo's first GW Transient Catalogue (GWTC-1)~\cite{LIGOScientific:2018mvr}; the spin distribution is noticeably bimodal, with initial BHs having low spins $\chi\lesssim0.2$ and final BHs have large spins $\chi\approx0.7$.
    (The spins of the initial BHs are not confidently measured in GWTC-1, so the spin values and uncertainties plotted here are merely heuristic and are estimated from the inspiralling binary's effective aligned spin $\chi_\mathrm{eff}$.)
    Green crosses show the SMBHs catalogued in Ref.~\cite{Brenneman:2011wz}.
    The grey region shows the expected parameter space for ``conventional'' PBHs formed from overdensities collapsing during radiation domination, including the effects of accretion as calculated in Ref.~\cite{DeLuca:2020bjf}.
    The magenta region shows a possible population of near-extremal PBHs formed during a period of early matter domination.
    The red region shows cusp-collapse PBHs, shaded according to their mass spectrum for $G\mu=10^{-11}$, cf.~Fig.~\ref{fig:mass-spectra-Gmu-1e-11}.
    All of the PBH populations are cut off at $m_*\approx3\times10^{-19}M_\odot$, due to evaporation through Hawking radiation.}
    \label{fig:regge-plane}
\end{figure}

Note that Eq.~\eqref{eq:mass-spectrum} includes only the energy density due to the rest mass of the PBHs; their kinetic energy at formation is larger by a factor of $G\mu$, though this will eventually be redshifted away.
This large kinetic energy will likely lead to interesting phenomenology and constraints which are not captured by traditional PBH analyses, as these generally assume the PBHs are formed with negligible peculiar velocities.
We plan to explore this in future work.

Eq.~\eqref{eq:mass-spectrum} does not include evaporation due to Hawking radiation, and is therefore only valid for masses greater than $m_*\approx5\times10^{14}\,\mathrm{g}\approx3\times10^{-19}M_\odot$, with PBHs lighter than this evaporating in less than a Hubble time~\cite{Carr:2020gox}.
Nonetheless, the form of $f(m)$ for masses below $m_*$ can be useful for deriving constraints on the overall mass spectrum due to evaporation effects.
For small masses in the radiation era, Eq.~\eqref{eq:mass-spectrum} approaches a time-independent power law
    \begin{equation}
    \label{eq:low-mass-spectrum}
        f(m)=f_*(m/m_*)^{-1/2},
    \end{equation}
    with $f_*$ a constant depending on $G\mu$ and on the network model.
The negative exponent means that the mass spectrum is dominated by very small PBHs, and that the strongest constraints on cusp collapse come from their evaporation.
In fact, Eq.~\eqref{eq:low-mass-spectrum} has the same power law as the mass spectrum resulting from the collapse of circular loops~\cite{MacGibbon:1997pu,James-Turner:2019ssu}, but with a different pre-factor.
In the circular collapse case, the pre-factor depends on the fraction of circular loops, which is unknown; in our case, the pre-factor depends only on $G\mu$, which we can therefore constrain directly.
Using the most up-to-date constraints from Ref.~\cite{James-Turner:2019ssu}, we find\footnote{%
The constraints in Ref.~\cite{James-Turner:2019ssu} are phrased in terms of a normalisation constant $c_\mathrm{string}$, with the CMB constraint giving $c_\mathrm{string}<2\times10^{-12}$.
This constraint can be translated to our mass spectrum~\eqref{eq:low-mass-spectrum} using $c_\mathrm{string}=2f_*\rho_\mathrm{CDM}/\rho_\mathrm{crit}$, where $\rho_\mathrm{crit}=3H_0^2/(8\uppi G)$ is the critical cosmological energy density.}
    \begin{equation}
    \label{eq:Gmu-bounds-CMB}
        G\mu<
        \begin{cases}
            6.0\times10^{-7} & \text{for model 1,}\\
            1.2\times10^{-6} & \text{for model 2,}\\
            9.3\times10^{-7} & \text{for model 3,}
        \end{cases}
    \end{equation}
    which in turn gives a constraint on the total fraction of DM made up by cusp-collapse PBHs,
    \begin{equation}
        \frac{\rho_\mathrm{PBH}}{\rho_\mathrm{CDM}}\equiv\int_{m_*}^\infty\dd{m}\frac{f(m)}{m}<
        \begin{cases}
            2.0\times10^{-10} & \text{for model 1,}\\
            7.4\times10^{-11} & \text{for model 2,}\\
            8.0\times10^{-10} & \text{for model 3.}
        \end{cases}
    \end{equation}
The constraint~\eqref{eq:Gmu-bounds-CMB} on $G\mu$ is comparable to those set by CMB analyses~\cite{Ade:2013xla,McEwen:2016aob}, and is almost independent of the network model, in stark contrast with, e.g., the LIGO constraints in Table~\ref{tab:Gmu-limits-LIGO}.
This constraint is set by the damping of small-scale CMB anisotropies due to PBHs decaying at recombination~\cite{Zhang:2007zzh,Carr:2009jm}, and is orders of magnitude stronger than the $\gamma$-ray constraint~\cite{James-Turner:2019ssu}.
It is likely to become more stringent with future CMB missions, and with similar analyses from upcoming 21cm experiments~\cite{Stocker:2018avm,Lucca:2019rxf}.

\section{A unique BH population}
\label{sec:populations}

\begin{figure}[t!]
    \includegraphics[width=0.48\textwidth]{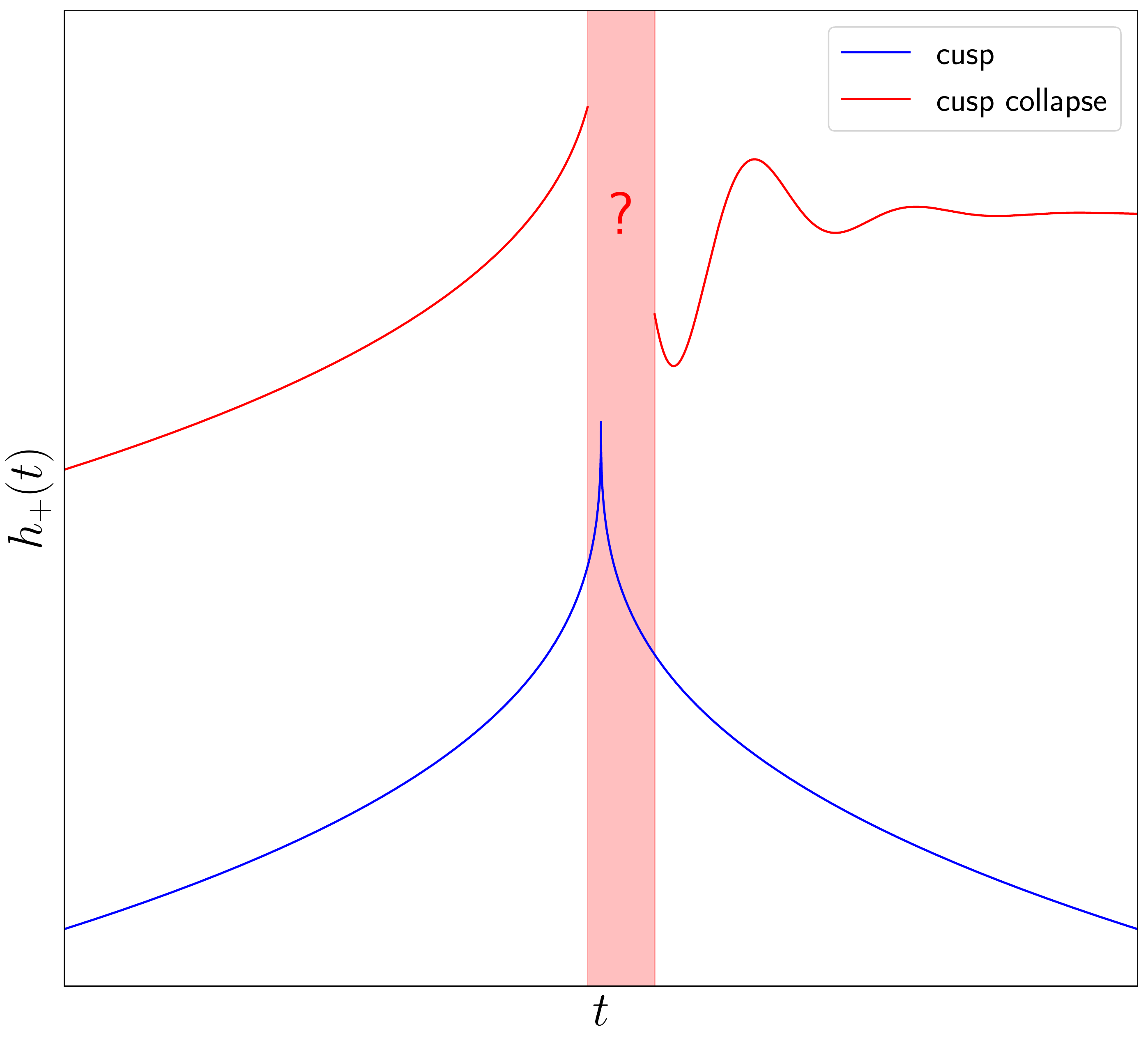}
    \caption{%
    A heuristic illustration of the different GW signals from collapsing and non-collapsing cusps.
    Shown in blue is the standard time-domain cusp waveform~\eqref{eq:cusp-waveform-td} from Ref.~\cite{Damour:2001bk}, which is symmetric around the peak.
    The cusp collapse waveform, in red, is truncated just before the peak, and eventually culminates in the QNM ringing of the final PBH.
    The uncertain period inbetween is denoted with a question mark.
    Note that the QNM frequency~\eqref{eq:qnm-omega} is much higher than depicted here, and that this figure is only for illustrative purposes.}
    \label{fig:waveform}
\end{figure}

We have shown that cusp-collapse PBHs are universally formed with dimensionless spins of $\chi=2/3$, regardless of the loop size $\ell$ or the string tension $G\mu$.\footnote{%
Note that this is the ``na\"{i}ve'' zero-radiation value, and that a fully general-relativistic calculation would likely give a different value for the final spin.
However, our argument in this section still holds, provided that the true value of $\chi$ is significantly larger than zero and less than unity.}
For sufficiently large rest masses $m\gg m_*$, this initial spin value is not affected by Hawking radiation, and survives to the present day~\cite{Arbey:2019jmj}.
This is interesting because it means that cusp-collapse PBHs occupy a unique region of the ``Regge plane''~\cite{Arvanitaki:2010sy,Berti:2015itd,Brito:2017zvb} (i.e., the BH mass-spin parameter space), as we illustrate in Fig~\ref{fig:regge-plane}.

Spins of $\chi\approx2/3$ are common amongst astrophysical BHs.
In particular, this is a very natural value for BHs formed from binary mergers like those observed by LIGO/Virgo~\cite{LIGOScientific:2018mvr}; the majority of such binaries are approximately equal-mass with small initial spins~\cite{LIGOScientific:2018jsj,Fishbach:2019bbm}, which correspond to final spins of $\chi\approx0.687$~\cite{Rezzolla:2007rz}.
SMBHs in active galactic nuclei are also observed to have large spins $\chi\gtrsim0.6$, due to accretion and prior mergers~\cite{Brenneman:2011wz}.
However, such astrophysical processes are unable to create subsolar-mass BHs, which dominate the cusp-collapse PBH mass spectrum for realistic values of $G\mu$, cf. Fig.~\ref{fig:mass-spectra-Gmu-1e-11}.

On the other hand, subsolar masses are generally possible in other PBH formation mechanisms, but these mechanisms are unable to generate spins $\chi\approx2/3$ like those resulting from cusp collapse.
``Conventional'' PBHs formed from collapsing overdensities during radiation domination are typically born with small spins of order $\chi\sim0.01$~\cite{Chiba:2017rvs,Mirbabayi:2019uph,DeLuca:2019buf}.
These initially low-spinning PBHs can acquire large spins through accretion, saturating the Thorne bound $\chi\approx0.998$~\cite{Thorne:1974ve}, but this process only takes place within a Hubble time if the PBHs in question are sufficiently massive, $m\gtrsim50\,M_\odot$~\cite{DeLuca:2020bjf}.
Subsolar-mass PBHs are extremely inefficient at accreting matter, and remain essentially non-spinning.

Subsolar-mass PBHs can have large spins if they form from collapsing overdensities during a hypothetical period of early matter domination~\cite{Polnarev:1986bi,Carr:2017edp,Harada:2017fjm,Kuhnel:2019zbc}, as radiation pressure is then unable to dissipate angular momentum during the collapse.
However, these PBHs are expected to have near-extremal spins $\chi\approx1$, which are easily distinguishable from the $\chi=2/3$ prediction of cusp collapse.

We therefore see that any observation of a subsolar-mass BH with a large (but non-extremal) spin $\chi\approx2/3$ would be incompatible with any of the other BH formation mechanisms mentioned here, and would be a ``smoking gun'' signature of cusp collapse, and of cosmic strings more generally.

\section{Consequences for GW searches}
\label{sec:gws}

\begin{figure}[t!]
    \includegraphics[width=0.48\textwidth]{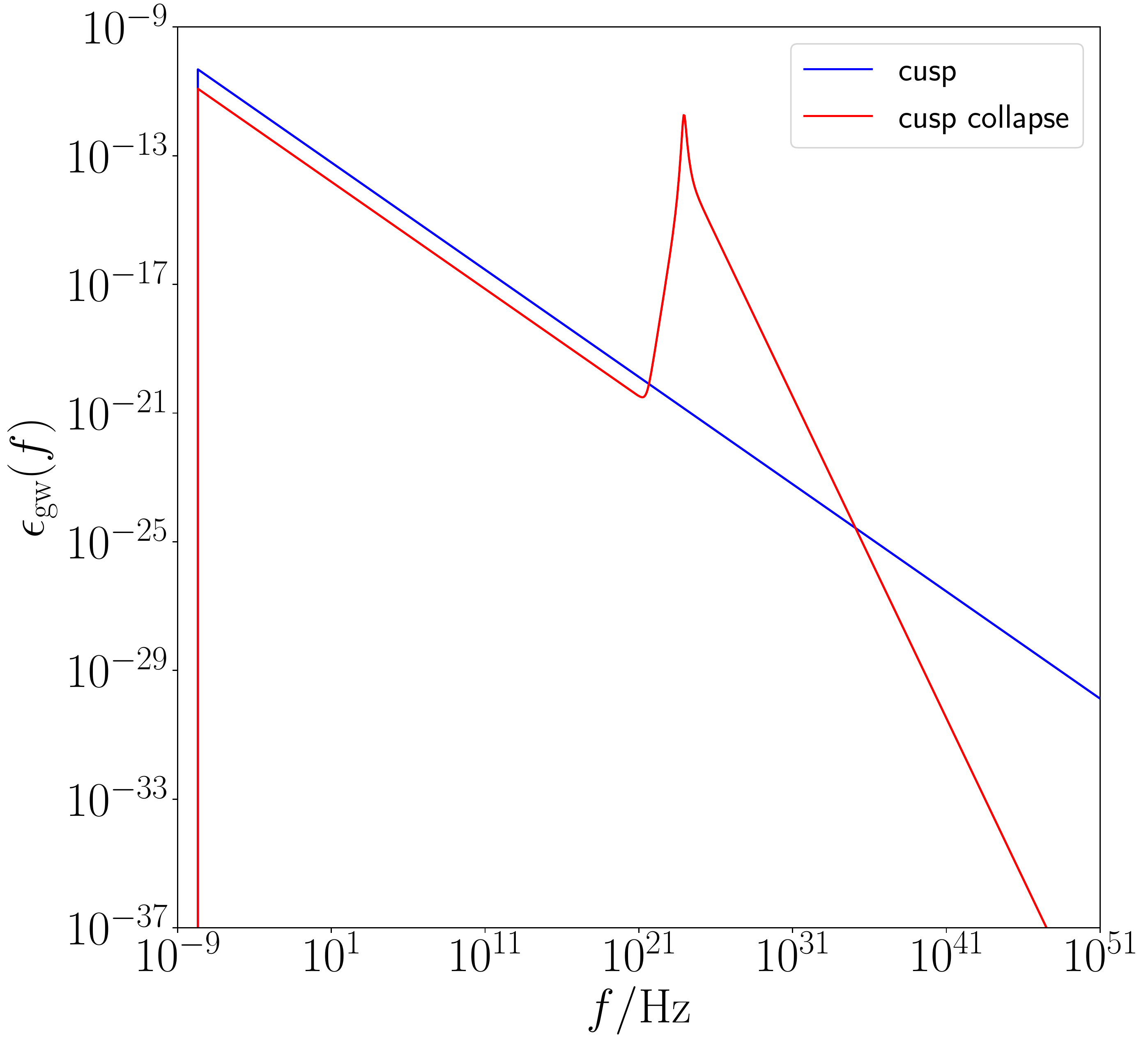}
    \caption{%
    The logarithmic one-sided GW energy spectrum as a fraction of the loop mass for cusps with and without collapse on a loop of length $\ell=1\,\mathrm{pc}$, as given by Eqs.~\eqref{eq:cusp-energy-spectrum} and~\eqref{eq:cusp-collapse-energy-spectrum} respectively.
    The cusp-collapse spectrum is smaller than that of a non-collapsing cusp by a factor of $1/4$ at low frequencies, has a strong peak at very high frequencies due to the QNM ringing of the PBH, and then decays like $\sim1/f$ (which is faster than the $\sim1/f^{1/3}$ cusp power law).
    There is a hard cutoff at low frequencies due to the fundamental mode of the loop, $f_0=2/\ell$.}
    \label{fig:cusp-collapse-gws}
\end{figure}

Cusps emit strong burst of GWs, and are promising potential sources for ground-based GW interferometers like LIGO/Virgo~\cite{Abbott:2017mem} and future space-based interferometers like LISA~\cite{Auclair:2019wcv}.
It is therefore important to understand how PBH formation affects the GW emission from the cusp.

The standard frequency-domain cusp waveform is derived in detail in Ref.~\cite{Damour:2001bk}.
While the low-frequency part of the waveform depends on the exact configuration of the loop, at high frequencies there is a universal power-law behaviour,
    \begin{equation}
    \label{eq:cusp-waveform-fd}
        \tilde{h}(f)\simeq A_f\frac{G\mu\ell^{2/3}}{r|f|^{4/3}},\qquad|f|\gg2/\ell,
    \end{equation}
    where $A_f\approx0.851$ is a numerical constant.
The emitted GWs are linearly polarised, and concentrated into a narrow beam of width $\theta_\mathrm{b}\simeq2^{2/3}3^{-1/6}(|f|\ell)^{-1/3}$.
In the time domain, the GW strain near the time of the peak $t_0$ can be approximated by
    \begin{equation}
    \label{eq:cusp-waveform-td}
        h(t)\simeq-A_t\frac{G\mu\ell^{2/3}}{r}|t-t_0|^{1/3},\qquad|t-t_0|\ll\ell/2,
    \end{equation}
    with $A_t\approx11.0$.

As a first approximation, we can model the effects of cusp collapse by truncating the standard time-domain waveform~\eqref{eq:cusp-waveform-td} at some time $t_\mathrm{PBH}<t_0$ when the horizon forms.
In the limit $t_\mathrm{PBH}\to t_0$ where the PBH forms at the peak of the cusp signal, the resulting frequency-domain waveform is exactly half of the standard one~\eqref{eq:cusp-waveform-fd}, with the other half corresponding to the truncated part of the signal at $t\ge t_0$.
Based on the discussion around Eq.~\eqref{eq:collapse-timescale}, we expect $t_0-t_\mathrm{PBH}\sim G\mu\ell\ll\ell/2$, so the waveform is truncated slightly before $t_0$, as shown in Fig.~\ref{fig:waveform}.
This leads to a loss of power at frequencies above $f_\mathrm{PBH}\sim1/(G\mu\ell)$.

A further contribution to the signal comes from the quasinormal ringing of the PBH.
We can describe this very approximately by including only the leading-order $(\ell,m,n)=(2,2,0)$ quasinormal mode (QNM), writing
    \begin{equation}
        h(t)\approx C\frac{Gm_\mathrm{PBH}}{r}\exp[\mathrm{i}(\omega t+\phi)-t/\tau],
    \end{equation}
    where $C$ and $\phi$ are unknown real constants.
Using the fitting formulae in Ref.~\cite{Berti:2005ys}, we take the real and imaginary parts of the $(2,2,0)$ QNM for a PBH with spin $\chi=2/3$ as
    \begin{equation}
        \omega\approx0.5214/(Gm_\mathrm{PBH}),\quad1/\tau\approx0.1715/(Gm_\mathrm{PBH}).
    \end{equation}
Since $Gm_\mathrm{PBH}\approx(G\mu)^3\ell$, we see that the ringdown signal is associated with extremely high frequencies,
    \begin{equation}
    \label{eq:qnm-omega}
        \omega\approx5.066\times10^{24}\,\mathrm{Hz}\times\qty(\frac{\ell}{\mathrm{pc}})^{-1}\qty(\frac{G\mu}{10^{-11}})^{-3}.
    \end{equation}

Our ignorance about the exact details of the collapse means that we cannot hope to construct an accurate phase-coherent waveform like those in Refs.~\cite{Helfer:2018qgv,Aurrekoetxea:2020tuw}.
However, by accounting for the truncation of the cusp signal and the PBH ringdown, we can obtain a reasonable approximation to the (logarithmic, one-sided) GW energy spectrum,
    \begin{align}
    \begin{split}
        \epsilon_\mathrm{gw}(f,\ell)&\equiv\frac{1}{\mu\ell}\dv{E_\mathrm{gw}}{(\ln f)}\\
        &=\frac{\uppi r^2f^3}{4G\mu\ell}\int_{S^2}\dd[2]{\vu*r}\qty(|\tilde{h}(f)|^2+|\tilde{h}(-f)|^2),
    \end{split}
    \end{align}
    which we have normalised with respect to the mass of the loop.
For the standard waveform~\eqref{eq:cusp-waveform-fd} this is
    \begin{equation}
    \label{eq:cusp-energy-spectrum}
        \epsilon_\mathrm{gw}(f,\ell)\simeq(2/3)^{1/3}\uppi^2A_f^2G\mu(f\ell)^{-1/3},
    \end{equation}
    where we have accounted for the beaming, which introduces a factor of $\uppi\theta_\mathrm{b}^2$.
For the cusp collapse case, we have instead
    \begin{align}
    \begin{split}
    \label{eq:cusp-collapse-energy-spectrum}
        \epsilon_\mathrm{gw}&\approx\frac{(2/3)^{1/3}}{4}\uppi^2A_f^2G\mu(f\ell)^{-1/3}\Theta(\omega-2\uppi f)\\
        &+\frac{C^2\ell}{4}(\uppi f)^3(G\mu)^5\qty[\mathcal{L}^2\qty(2\uppi f;\omega,\tfrac{1}{\tau})+\mathcal{L}^2\qty(2\uppi f;-\omega,\tfrac{1}{\tau})],
    \end{split}
    \end{align}
    where the first part is reduced by a factor $1/4$ compared to Eq.~\eqref{eq:cusp-energy-spectrum} (due to a factor $1/2$ in each power of the strain) and is truncated at the QNM frequency~\eqref{eq:qnm-omega} by the step function $\Theta$, while the second part is the ringdown contribution, written in terms of the Lorentzian
    \begin{equation}
        \mathcal{L}(x;x_0,\gamma)\equiv\frac{\gamma/\uppi}{\gamma^2+(x-x_0)^2}.
    \end{equation}

We can fix the constant $C$ by setting the total energy radiated by the ringdown term equal to $G\mu\epsilon_M$, where $\epsilon_M$ is the collapse radiation efficiency introduced in Eq.~\eqref{eq:radiation-efficiency}, and the factor $G\mu$ translates between the loop's mass and the PBH's relativistic mass (i.e. rest mass plus kinetic energy), with $\epsilon_M$ defined as a fraction of the latter.
This gives
    \begin{equation}
        C=\sqrt{\frac{64\uppi\epsilon_M}{1+\omega^2\tau^2}\frac{\tau/Gm_\mathrm{PBH}}{G\mu}}\approx10.70\times(\epsilon_M/G\mu)^{1/2}.
    \end{equation}
We assume a value of $\epsilon_M=10\%$, which is consistent with the upper bounds~\eqref{eq:spin-bound-33} and~\eqref{eq:spin-bound-18}, and is comparable to the mass-radiation fraction found in numerical simulations of other ultrarelativistic, strong-gravity phenomena~\cite{Sperhake:2008ga,East:2012mb}.
The resulting GW energy spectrum is shown in Fig.~\ref{fig:cusp-collapse-gws}.
The total fraction of the loop's mass radiated by the cusp is approximately
    \begin{equation}
    \label{eq:gw-radiation-fraction}
        \int_{2/\ell}^\infty\frac{\dd{f}}{f}\epsilon_\mathrm{gw}\approx
        \begin{cases}
            14.9\,G\mu & \text{cusp},\\
            (3.71+\epsilon_M)\,G\mu & \text{cusp collapse},
        \end{cases}
    \end{equation}
    which shows that the radiation from collapsing cusps is comparable to, but strictly less than, that from non-collapsing cusps.

\begin{figure}[t!]
    \includegraphics[width=0.48\textwidth]{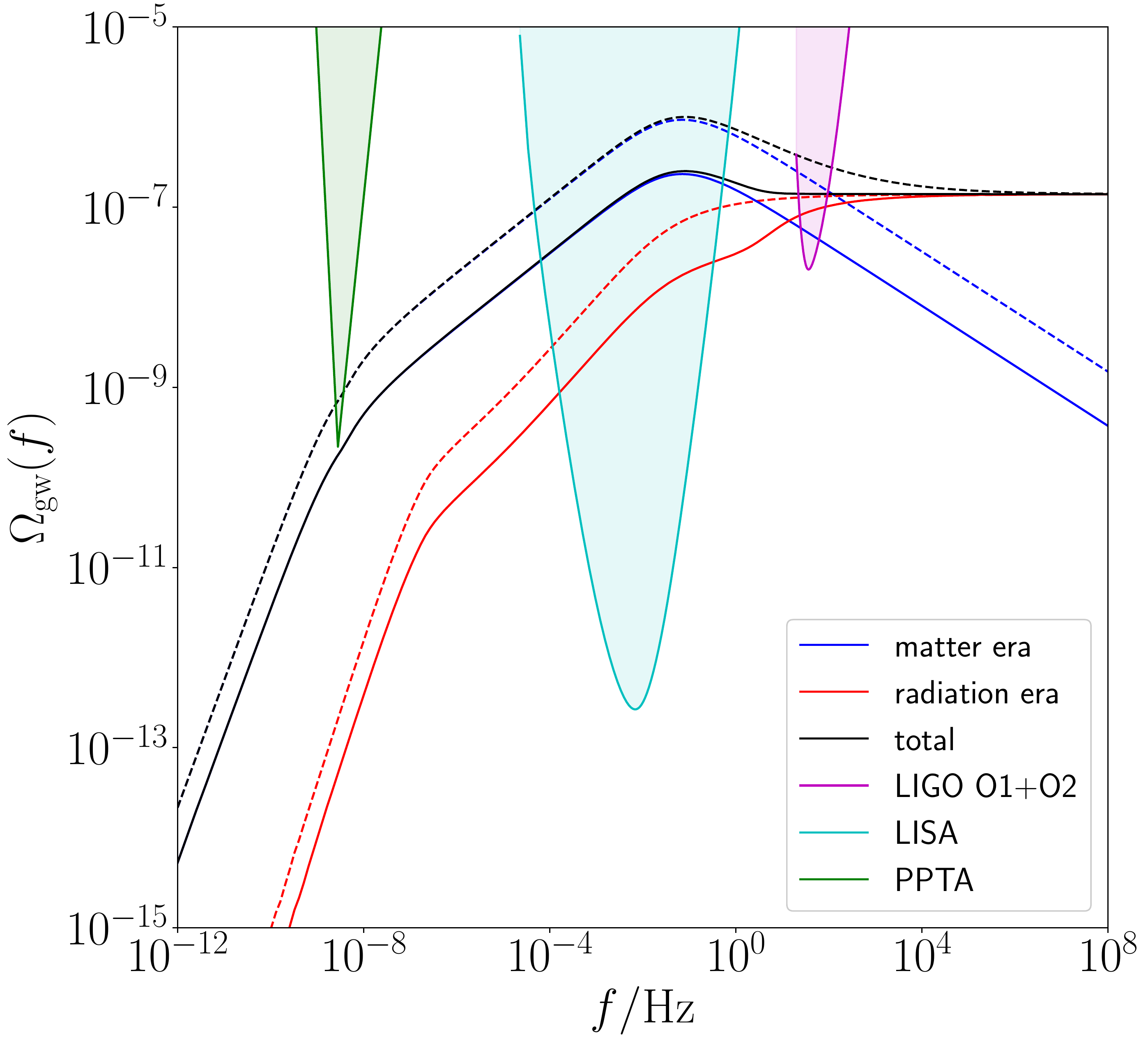}
    \caption{%
    The SGWB spectrum from cusps on cosmic string loops.
    Solid lines include the effects of cusp collapse using Eq.~\eqref{eq:cusp-collapse-energy-spectrum}, while dotted lines correspond to the standard case without collapse~\eqref{eq:cusp-energy-spectrum}.
    The magenta curve shows the power-law-integrated (PI) sensitivity curve~\cite{Thrane:2013oya} from the LIGO O1+O2 isotropic stochastic search~\cite{LIGOScientific:2019vic}, which is publicly available at Ref.~\cite{LIGOScientific:2019PI}.
    The green curve shows the Parkes Pulsar Timing Array (PPTA) PI curve~\cite{Shannon:2015ect,Verbiest:2016vem}, calculating using the code from Ref.~\cite{Thrane:2013oya}, which is publicly available at Ref.~\cite{Thrane:2013PI}.
    The cyan curve shows the projected LISA power-law-integrated sensitivity curve, as described in Refs.~\cite{Caprini:2019pxz,Smith:2019wny}.
    We use model 3 of the loop network~\cite{Ringeval:2005kr,Lorenz:2010sm} with $G\mu=3\times10^{-11}$, illustrating how the PPTA bound is weakened due to cusp collapse.
    At high frequencies the spectra with and without cusp collapse become identical; the frequency at which this changeover occurs decreases for smaller values of $G\mu$, meaning that the LIGO bound on model 3 is the same in both cases.}
    \label{fig:Omega_gw}
\end{figure}

The combined GW emission from many loops throughout cosmic history gives rise to a stochastic GW background (SGWB)~\cite{Vilenkin:1981bx,Hogan:1984is,Vachaspati:1984gt,Accetta:1988bg,Bennett:1990ry,Caldwell:1991jj,Allen:1996vm,Maggiore:1999vm,Damour:2000wa,Damour:2001bk,Damour:2004kw,Siemens:2006yp,DePies:2007bm,Olmez:2010bi,Binetruy:2012ze,Sanidas:2012ee,Kuroyanagi:2012wm,Kuroyanagi:2012jf,Sanidas:2012tf,Sousa:2013aaa,Blanco-Pillado:2013qja,Henrot-Versille:2014jua,Blanco-Pillado:2017rnf,Blanco-Pillado:2017oxo,Ringeval:2017eww,Cui:2017ufi,Abbott:2017mem,Caprini:2018mtu,Jenkins:2018nty,Cui:2018rwi,Christensen:2018iqi,LIGOScientific:2019vic,Auclair:2019wcv,Auclair:2019jip}.
The intensity of the SGWB is usually expressed as a function of frequency in terms of the density parameter,
    \begin{equation}
        \Omega_\mathrm{gw}(f)\equiv\frac{1}{\rho_\mathrm{crit}}\dv{\rho_\mathrm{gw}}{(\ln f)},
    \end{equation}
    which for cosmic string cusps is given by
    \begin{equation}
        \Omega_\mathrm{gw}(f)=\frac{16\uppi}{3H_0^2}G\mu N_\mathrm{cusp}\int\dd{t}\frac{a^4}{t^3}\int\dd{\gamma}\mathcal{F}(\gamma)\epsilon_\mathrm{gw}(f/a,\gamma t).
    \end{equation}
We account for cusp collapse by integrating over the GW energy spectrum~\eqref{eq:cusp-collapse-energy-spectrum}.
A representative example of the resulting SGWB spectrum is shown in Fig.~\ref{fig:Omega_gw}.
At low frequencies $\Omega_\mathrm{gw}$ is reduced by a factor of $1/4$ compared to the standard spectrum, which relaxes the constraints on $G\mu$ coming from LIGO~\cite{LIGOScientific:2019vic} and from PTAs~\cite{Shannon:2015ect,Lasky:2015lej,Verbiest:2016vem,Blanco-Pillado:2017rnf}.
At high frequencies this factor $1/4$ difference vanishes, as the signal is dominated by loops which are too small to undergo cusp collapse.
The changeover between these two regimes depends on the value of $G\mu$, as this sets the size of the smallest loops that can undergo cusp collapse through Eq.~\eqref{eq:ell-min-cusp-collapse}; for smaller values of $G\mu$ the changeover happens at lower frequencies.
At very high frequencies the QNM emission from PBHs forming in the matter era gives rise to a strong peak, but this is dwarfed by the radiation-era plateau, making it unobservable.

In Tables~\ref{tab:Gmu-limits-LIGO},~\ref{tab:Gmu-limits-PPTA}, and~\ref{tab:Gmu-limits-LISA} we present updated bounds on $G\mu$ from SGWB searches with LIGO and with PTAs, as well as forecasts for LISA~\cite{Audley:2017drz}, accounting for the modified GW emission due to cusp collapse.
We include GW emission from kinks and kink-kink collisions, which occur $N_\mathrm{kink}=4.53$ and $N_\text{kink-kink}=N_\mathrm{kink}^2/4$ times per oscillation period, respectively.\footnote{%
This value of $N_\mathrm{kink}$ is required when $N_\mathrm{cusp}=1$ to ensure the total GW power is consistent with that used in the loop network models---see Eqs.~(A.9)--(A.11) of Ref.~\cite{Auclair:2019wcv} for more details.
The expression for $N_\text{kink-kink}$ is due to there being equal numbers of left- and right-moving kinks, with each left-mover colliding with every right-mover once per loop oscillation, and vice versa.}
The reduction in GW power at low frequencies due to cusp collapse relaxes the LIGO bounds on models 1 and 2 by an order of magnitude; model 3 is unaffected however, as the existing bound of $G\mu\lesssim10^{-14}$ is small enough that the reduction in power only happens at frequencies below the LIGO band.
Similarly, the LISA bounds are so strong that there is no reduction in power in the LISA band for the corresponding values of $G\mu$, and the resulting constraints are identical with or without cusp collapse.
The PTA bounds are weakened by a factor of $\approx2$ for all three network models, with the bound for model 3 illustrated in Fig.~\ref{fig:Omega_gw}.

\begin{table}[t!]
    \caption{\label{tab:Gmu-limits-LIGO}%
    Upper limits (95\% confidence) on $G\mu$ from the LIGO O1+O2 isotropic stochastic search~\cite{LIGOScientific:2019vic,LIGOScientific:2019PI} for each of the three loop network models, with and without cusp collapse.
    These limits are stronger than those from the LIGO/Virgo search for resolvable GW bursts from cosmic strings~\cite{Abbott:2009rr,Aasi:2013vna,Abbott:2017mem}.
    We derive the limits by comparing the O1+O2 PI curve~\cite{Thrane:2013oya} from Ref.~\cite{LIGOScientific:2019PI} with our calculated spectra, including kinks and kink-kink collisions as described in the main text.
    This prescription is slightly different to that used in Ref.~\cite{LIGOScientific:2019vic}, with a greater number of kinks and kink-kink collisions, which is why the limits for models 2 and 3 are slightly stronger than those reported there.
    No limits for model 1 were reported in Ref.~\cite{LIGOScientific:2019vic}, due to that model being disfavoured by numerical simulations, as mentioned previously.}
    \begin{ruledtabular}
    \begin{tabular}{c | c c}
        $G\mu$ bounds & without cusp collapse & with cusp collapse \\
        \hline\hline
        model 1 & $1.6\times10^{-9}$ & $1.9\times10^{-8}$ \\
        \hline
        model 2 & $1.3\times10^{-7}$ & $1.6\times10^{-6}$ \\
        \hline
        model 3 & $2.0\times10^{-14}$ & $2.0\times10^{-14}$
    \end{tabular}
    \end{ruledtabular}
\end{table}

\begin{table}[t!]
    \caption{\label{tab:Gmu-limits-PPTA}%
    Upper limits (95\% confidence) on $G\mu$ from the PPTA~\cite{Shannon:2015ect,Lasky:2015lej,Blanco-Pillado:2017rnf} for each of the three loop network models, with and without cusp collapse.
    These limits are stronger than those from NANOGrav and from the European Pulsar Timing Array~\cite{Verbiest:2016vem}.}
    \begin{ruledtabular}
    \begin{tabular}{c | c c}
        $G\mu$ bounds & without cusp collapse & with cusp collapse \\
        \hline\hline
        model 1 & $8.0\times10^{-12}$ & $1.7\times10^{-11}$ \\
        \hline
        model 2 & $3.1\times10^{-11}$ & $8.1\times10^{-11}$ \\
        \hline
        model 3 & $1.4\times10^{-11}$ & $3.0\times10^{-11}$
    \end{tabular}
    \end{ruledtabular}
\end{table}

\begin{table}[t!]
    \caption{\label{tab:Gmu-limits-LISA}%
    Forecast upper limits (95\% confidence) on $G\mu$ from LISA.
    We use the LISA power-law-integrated sensitivity curve~\cite{Caprini:2019pxz,Smith:2019wny}, and assume a 75\% observing duty cycle over the nominal 4-year mission, giving a 3-year dataset.
    For all three network models, $G\mu$ is so small that there is no reduction in power in the LISA band, and the resulting constraints are identical with or without cusp collapse.}
    \begin{ruledtabular}
    \begin{tabular}{c | c c}
        $G\mu$ bounds & without cusp collapse & with cusp collapse \\
        \hline\hline
        model 1 & $6.9\times10^{-18}$ & $6.9\times10^{-18}$ \\
        \hline
        model 2 & $3.0\times10^{-17}$ & $3.0\times10^{-17}$ \\
        \hline
        model 3 & $2.9\times10^{-17}$ & $2.9\times10^{-17}$
    \end{tabular}
    \end{ruledtabular}
\end{table}

\section{Summary and conclusion}
\label{sec:summary}

We have shown that cusps (and pseudocusps) on cosmic string loops satisfy the hoop condition~\eqref{eq:hoop-condition}, and are therefore predicted to undergo gravitational collapse to form PBHs.\footnote{%
Of course, while extremely successful and useful, the hoop conjecture is only a conjecture.
In the (seemingly unlikely) case that cusps do not collapse in general relativity, they would constitute the first known counterexample to the hoop conjecture, which would itself be of fundamental importance.}
Since cusps are generic, forming typically once per loop oscillation, this implies that the rate of PBH production from cosmic strings has been drastically underestimated in the literature, where for decades it has been assumed that the main mechanism for PBH formation from cosmic strings is through the collapse of very rare (quasi)circular loops~\cite{Hawking:1987bn,Polnarev:1988dh,Hawking:1990tx,Garriga:1992nm,Caldwell:1993kv,Garriga:1993gj,Caldwell:1995fu,MacGibbon:1997pu,Helfer:2018qgv,James-Turner:2019ssu,Aurrekoetxea:2020tuw}.
While our analysis is based on the flat-space equations of motion~\eqref{eq:EoM}, we have argued that gravitational backreaction acts far too slowly to prevent the collapse.

We have argued that, due to the unstable configuration of the parent loop near the horizon, large cusp-collapse PBHs are likely to be rapidly cut off from the loop network by string self-intersection.
The majority of cusp-collapse PBHs, however, are extremely small, and evaporate on short timescales.
These evaporating PBHs are constrained by their damping of small-scale CMB anisotropies~\cite{Zhang:2007zzh,Carr:2009jm,James-Turner:2019ssu}, which leads to a nearly model-independent bound on the string tension, $G\mu\lesssim10^{-6}$.
This in turn implies that cusp-collapse PBHs can only make up a small fraction of the DM.

By calculating the angular momentum of the string segment captured behind the horizon, we have shown that cusp-collapse PBHs are highly spinning, with dimensionless spin parameter equal to two-thirds of the extremal Kerr value, $\chi=2/3$.
This spin is a universal property of the formation mechanism, and is independent of the loop size $\ell$ and string tension $G\mu$.
To the best of our knowledge, cusp collapse is the only known primordial or astrophysical mechanism for generating subsolar-mass BHs with large but sub-extremal spins.
The observation of such a BH would therefore be a ``smoking gun'' signal of cusp collapse, and of cosmic strings more generally.

In the absence of exact solutions for the collapse, we have developed a simple approximation for the expected GW signal, based on the standard cusp waveform of Ref.~\cite{Damour:2001bk}.
At low frequencies, the radiated GW energy spectrum is reduced by a factor of $1/4$ compared to the standard cusp waveform, due to the truncation of the signal at, or just before, the peak of the cusp.
At very high frequencies, there is a strong contribution due to the QNM ringing of the newly-formed PBH.
Integrating this GW emission over the cosmic string loop distribution, we have obtained updated predictions for the SGWB spectrum.
The reduction of the SGWB intensity at frequencies probed by LIGO/Virgo and PTAs relaxes existing constraints on $G\mu$ by as much as an order of magnitude, depending on the GW frequency band and the loop network model.
However, for sufficiently small values of $G\mu$ the SGWB spectrum is unchanged at observable frequencies, and the corresponding constraints are unaffected---this is the case for LISA, as well as for the LIGO constraint on model 3.

The results presented here constitute a significant development in our understanding of the dynamics of cosmic string loops, and of the observational and cosmological consequences of loop networks.
This opens up several avenues for further research.
In particular, it seems important to develop a full description of the collapse, going beyond the flat-space equations of motion for the loop.
As in the case of circular loop collapse, fully general-relativistic calculations---whether analytical~\cite{Hawking:1990tx} or numerical~\cite{Helfer:2018qgv,Aurrekoetxea:2020tuw}---will be necessary to better understand the GW emission and the final properties of the PBHs.
This is likely to be extremely challenging however, due to the velocities involved, the lack of isometries, and the huge ratio of scales (see Fig.~\ref{fig:scales}).
The fact that cusp-collapse PBHs are born with ultrarelativistic velocities will undoubtedly give rise to some interesting phenomenology, and may allow us to place new constraints on their abundance, aside from those in the standard PBH literature.
It would also be very interesting to calculate the merger rate of cusp-collapse PBH binaries, as well as the corresponding SGWB spectrum~\cite{Mandic:2016lcn,Bartolo:2016ami,Clesse:2016ajp,Wang:2016ana,Raidal:2017mfl,Raidal:2018bbj}, as consistency with LIGO/Virgo observations (in particular the subsolar-mass search~\cite{Abbott:2005pf,Abbott:2018oah,Magee:2018opb,Authors:2019qbw}) would provide another independent constraint on the string tension; we leave this for future work.

\begin{acknowledgments}
    We acknowledge valuable comments and feedback from Josu Aurrekoetxea, Thomas Helfer, and Eugene Lim.
    A.~C.~J. is supported by King's College London through a Graduate Teaching Scholarship.
    M.~S. is supported in part by the Science and Technology Facility Council (STFC), United Kingdom, under the research grant No. ST/P000258/1.
    This is LIGO document number P2000218.
\end{acknowledgments}

\bibliography{pbh-from-cs}
\end{document}